\DocumentMetadata{}
\documentclass[sigconf]{acmart}

\usepackage{algorithm,algorithmic}
\usepackage{enumitem}

\usepackage{hyperref}
\usepackage{hyperxmp}
\usepackage{graphicx}
\usepackage{xspace}
\usepackage{xcolor}
\usepackage{subfig}
\usepackage{booktabs}
\usepackage{boxedminipage}
\usepackage{amsthm}
\usepackage{amsmath,amsfonts}
\usepackage{tabularray}

\usepackage{doi}
\usepackage[subtle]{savetrees}
\usepackage[font=footnotesize,labelfont=bf]{caption}

\hypersetup{
    colorlinks,
    linkcolor={red!50!black},
    citecolor={blue!50!black},
    urlcolor={blue!80!black}
}

\def\BibTeX{{\rm B\kern-.05em{\sc i\kern-.025em b}\kern-.08em
    T\kern-.1667em\lower.7ex\hbox{E}\kern-.125emX}}

\newtheorem{theorem}{Theorem}

\newtheorem{lemma}[theorem]{Lemma}

\newtheorem{observation}[theorem]{Observation}
\newtheorem{definition}[theorem]{Definition}

\newcommand{\calF}{{\mathcal F}}
\newcommand{\calS}{{\mathcal S}}
\newcommand{\sS}{{\calS^{\star}}}

\newcommand{\bigO}{{\mathcal O}}
\newcommand{\calM}{{\mathcal M}}

\newcommand{\I}{{\mathcal I}}
\newcommand{\hI}{\hat{\I}}
\newcommand{\Is}{{\I^{\star}}}
\newcommand{\Js}{{J^{\star}}}
\newcommand{\JA}{{J^{Alg}}}
\newcommand{\SA}{{S^{Alg}}}
\newcommand{\tJ}{\tilde{J}}
\newcommand{\tI}{\tilde{I}}
\newcommand{\tS}{\tilde{\calS}}
\newcommand{\bJ}{\bar{J}}

\newcommand{\tT}{{\mathcal T}}

\newcommand{\dpmss}{\textsc {DPMSS}\xspace}
\newcommand{\dpmssf}{\textsc {DPMSSF}\xspace }
\newcommand{\maxpft}{\textsc {Max-Profit} }
\newcommand{\lsdsf}{\textsc {LSDSF}\xspace}
\newcommand{\lsds}{\textsc {LSDS}\xspace}
\newcommand*\wbar[1]{%
   \hbox{%
     \vbox{%
       \hrule height 0.5pt % The actual bar
       \kern0.3ex%         % Distance between bar and symbol
       \hbox{%
         \kern-0.1em%      % Shortening on the left side
         \ensuremath{#1}%
         \kern-0.1em%      % Shortening on the right side
       }%
     }%
   }%
}

\renewcommand\footnotetextcopyrightpermission[1]{}

\begin{document}
\title{A Deadline-Aware Scheduler for Smart Factory using WiFi 6}
\author{Mohit Jain}
\email{mohit20221@iiitd.ac.in}
\affiliation{%
  \institution{IIIT Delhi, India}
  \country{} 
}
\author{Anis Mishra}
\email{anis20026@iiitd.ac.in}
\affiliation{%
  \institution{IIIT Delhi, India}
  \country{} 
}
\author{Syamantak Das}
\email{syamantak@iiitd.ac.in}
\affiliation{%
  \institution{IIIT Delhi, India}
  \country{} 
}
\author{Andreas Wiese}
\email{andreas.wiese@tum.de}
\affiliation{%
  \institution{TU Munich, Germany}
  \country{} 
}
\author{Arani Bhattacharya}
\email{arani@iiitd.ac.in}
\affiliation{%
  \institution{IIIT Delhi, India}
  \country{} 
}
\author{Mukulika Maity}
\email{mukulika@iiitd.ac.in}
\affiliation{
  \institution{IIIT Delhi, India}
  \country{} 
}

\begin{abstract}
A key strategy for making production in factories more efficient is to collect data about the functioning of machines, and dynamically adapt their working.
Such smart factories have data packets with a mix of stringent and non-stringent deadlines with varying levels of importance that need to be delivered via a wireless network.
However, the scheduling of packets in the wireless network is crucial to  satisfy the deadlines.
In this work, we propose a technique of utilizing IEEE 802.11ax, popularly known as WiFi 6, for such applications.
IEEE 802.11ax has a few unique characteristics, such as specific configurations of dividing the channels into resource units (RU) for packet transmission and synchronized parallel transmissions.
We model the problem of scheduling packets by assigning profit to each packet and then maximizing the sum of profits.
We first show that this problem is strongly NP-Hard, and then propose an approximation algorithm with a 12-approximate algorithm.
Our approximation algorithm uses a variant of local search to associate the right RU configuration to each packet and identify the duration of each parallel transmission.
Finally, we extensively simulate different scenarios to show that our algorithm works better than other benchmarks.

\end{abstract}

\maketitle

\section{Introduction}
\label{sec:into}
The increasing demand for more efficient and more precise manufacturing has led to the demand for more intelligent or ``smart factory" \cite{smart-factory,automation-factory}.
With manufacturing processes getting more complex, wired connectivity is gradually getting replaced by wireless technologies to avoid clutter on the factory floor, reduce maintenance costs and enable easier planning~\cite{automation-factory,flexible_iot_ieee_2020}.

\begin{figure}
    \centering
    \includegraphics[width=0.35\textwidth]{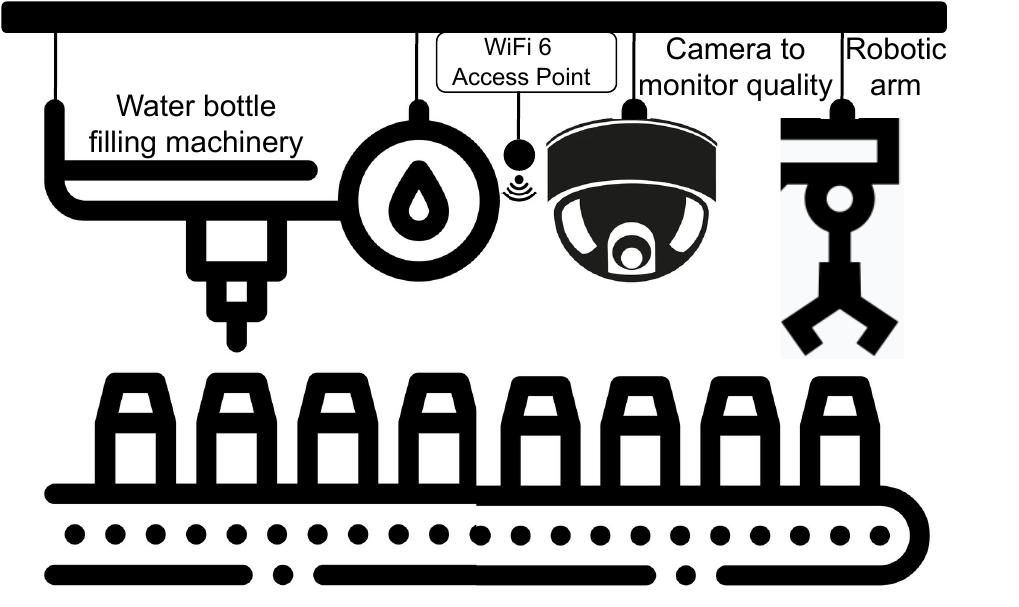}
    \vspace{-0.2in}
    \caption{An illustrative example of a bottle-filling factory being controlled over WiFi. The bottle-filling equipment and the robotic arm have much stringent deadlines and higher profit in case of failure than the camera monitoring quality.}
    \label{fig:illustration}
   \vspace{-0.45cm}
\end{figure}

\noindent \textbf{Background on Smart Factory:} A smart factory is able to dynamically adapt the operation of machines in response to changes in the condition of machines, thus leading to more predictable and efficient production.
Such a smart factory would collect data deployed from different sensors within the premises, and utilize them to schedule the operation rate of different machines.
Thus, different machines are connected by a network, and a controller needs to decide how to schedule and operate them.
A key challenge of such smart factory settings is that they handle a mix of critical and non-critical tasks. For example, Fig.~\ref{fig:illustration} shows a bottle-filling factory setting being controlled by WiFi. Here the \textit{water bottle filling machinery} and \textit{robotic arm} are very crucial to the operation of the factory. They have a very stringent deadline  as well compared to the camera who is monitoring quality.
%Table.~\ref{tab:factory-applications} lists some sample applications and their characteristics seen in factory scenarios~\cite{flexible_iot_ieee_2020}.
Here, critical operations %like human safety monitoring such as proximity near a machine or equipment control 
need to coexist with basic quality supervision and factory resource management.
Missed deadlines of the most critical applications could potentially lead to the shutdown of factory operations or fatal accidents. On the other hand, such missed deadlines for non-critical applications might only lead to a temporary slowdown of production. 
Such settings require strong performance guarantees on latency from the wireless network that is used to connect the factory equipment~\cite{flexible_iot_ieee_2020}.
Thus, wireless network technologies that are used need to ensure that they provide such performance guarantees.

A technique typically used for such applications is to first implement such protocols in the digital twin of a factory \cite{digital-twin-1}. Once the digital twin shows acceptable performance, the wireless technology is deployed. This step of integrating the simulation with digital twin leads to a better understanding of bandwidth requirements, the number of packets that miss their latency, and consequences of such misses. Thus, simulation of algorithms used to schedule packets in a specific technology is crucial to understand its suitability for deployment.

\iffalse
\begin{table}[t]
    \footnotesize
    \centering
    \begin{tabular}{|p{2.3cm}|l|p{1.6cm}|l|p{1cm}|}
    \hline
      \textbf{Application} & \textbf{Data (Bytes)} & \textbf{Comm. Rate} & \textbf{Delay Tol.} \\
     \hline
        \multicolumn{4}{|c|}{Equipment Control} \\
     \hline
      Control liquid injection & $64$ &  Once per $1$ min & $100$ ms \\
     \hline
      Automated Guided Vehicle (AGV) Control & $100$  &  Once per $1$ min & $100$ ms \\
     \hline
        Bottle filling & $400$  &  Once per $1$ ms & $0.5$ ms \\
     \hline
      Warehouse &  $10$  & Once per $2$ ms & $1$ ms \\
     \hline
        \multicolumn{4}{|c|}{Quality Supervision} \\
     \hline
        Monitoring of equipment & A few hundreds & Once per $1$ s & $1$ s \\
     \hline
        Counting number of wrench operations & $64$ & Once per $1$ min & $100$ ms \\
     \hline
        Detect defect state & $500$ & Once per $100$ ms & $500$ ms \\
     \hline
        \multicolumn{4}{|c|}{Factory Resource Management} \\
     \hline
        Movement analysis & A few tens & Twice per $1$ s & Few seconds \\
     \hline
        Work record & $100$ & Once per $1$ min & $1$ s \\
     \hline
        \multicolumn{4}{|c|}{Human Safety} \\
     \hline
        Detect entry in the proximity of a machine & $10$ - $30$ & Once per $1$ - $10$ ms & $2$ - $20$ ms \\    
     \hline
    \end{tabular} 
    \caption{Communication Requirements for various wireless applications in  factory settings}
    \vspace{-0.4cm}
    \label{tab:factory-applications}
\end{table}
\fi
\noindent \textbf{WiFi for Smart Factory:} Although WiFi is one of the most widely used wireless network technology, it was traditionally not used in factory settings because of its decentralized nature.
This decentralized contention-based channel access scheme of WiFi led to the possibility of collisions and thus a lack of any real-time guarantees.
Furthermore, WiFi was traditionally designed for a limited number of users/devices in a small area, thus limiting its scope in factory settings.

\noindent \textbf{What makes WiFi 6 Suitable for Smart Factory?} Compared to traditional WiFi, recent standards of WiFi such as IEEE 802.11ax, popularly known as WiFi 6, have made a number of changes that enable its use in factory settings. The first change is to enable its use in dense scenarios by replacing Orthogonal Frequency-Division Multiplexing (OFDM) based single-user transmissions with Orthogonal Frequency-Division Multiple Access (OFDMA) based multi-user transmissions.
OFDMA allows parallel transmissions by breaking the channel into orthogonal sets of sub-carriers (or tones) called resource units (RU) and assigning them to multiple users for each transmission~\cite{11ax-standard}.
This feature enables support for a smaller but more number of packets per unit time, which is commonly observed in factory settings \cite{flexible_iot_ieee_2020}. In addition, with the OFDMA based transmission, WiFi enables centralized control where the access point schedules both downlink and uplink transmissions. Furthermore, WiFi 6 also introduces more deterministic and centralized channel access scheme whereby using a specific procedure called Multi-User Enhanced Distributed Channel Access (MU-EDCA), it is possible to suspend contention for channel access by setting the EDCA timer~\cite{muedca-impact} for a specific duration. 
The stations switch their contention parameters to MU EDCA parameters for a duration specified by the timer. Hence, with MU-EDCA  distributed contention-based channel access scheme is completely replaced by a centralized scheme.
These changes enable WiFi 6 to be used in industrial settings \cite{seno_enhancing_2017,white-paper}.

In addition, utilizing WiFi 6 in smart factories also has a number of other advantages over cellular networks: (1) WiFi 6 APs are already commercially available and are cost-effective, making it easy to use in any factory. In contrast, using 5G requires acquiring of license from the government or approaching a telecom operator which provides specific URLLC service, which today are only available via commercial agreements and are not easily accessible to small factories~\cite{tsn-survey}. (2) Practical evaluations using 5G and WiFi suggested that both can provide latencies required for smart factory~\cite{5g-wifi-evaluation}, with WiFi providing occasionally better latency than 5G.

\noindent \textbf{Challenges:} In this paper, we utilize the paradigm shift towards centralized channel access supported by 802.11ax to design a scheduling framework for deadline-based flexible factory settings that involve uplink traffic. Although 802.11ax enables support for such factory settings, the actual scheduling of packets by the access point is still crucial to ensure that the factory is both safe and efficient.
Such scheduling must prioritize the delivery of the most critical packets within the deadline, while ensuring that the non-critical ones are not ignored.
The protocol itself also has its own characteristics not found in other technologies, such as 5G.
For example, (1) the standard provides a specific way of splitting the bandwidth into RUs (resource units), (2) one RU can be allocated to at most one user and vice-versa, and (3) the starting and ending time of the transmissions should be synchronized. These characteristics of 802.11ax make the scheduling of packets a challenge, as they do not allow us to utilize any algorithm proposed by prior works in scheduling (details in \S II).

\noindent \textbf{Our Approach:} Our solution utilizes known packet arrival patterns and deadline requirements for the applications. Further it defines a `profit' for each job that indicates the relative criticality with more critical packets having higher profit. For example, in case of the bottle filling factory (shown in Fig.~\ref{fig:illustration}), \textit{water bottle filling machinery} and \textit{robotic arm} are the critical applications hence they have higher profit compared to \textit{Camera to monitor quality} application. Such presence of critical applications co-existing with less critical ones is frequently seen in modern industrial systems \cite{mixed-critical}. The algorithm schedules packet transmissions provided by the access point via a trigger frame. The objective of the scheduler is to maximize the profit of packets that finish transmission on or before the respective deadline (\S~\ref{subsec:model}).
 
We reduce the above problem to a novel variant of classical parallel machine scheduling which we call \emph{\underline{D}eadline-aware \underline{P}arallel \underline{M}achines \underline{S}cheduling with \underline {S}ynchronized start (\dpmss)}. The two crucial constraints of scheduling jobs in `batches' and bandwidth restrictions on the RUs to be used pose new challenges over existing techniques for deadline-aware scheduling. We argue that our problem is strongly NP-hard even in the simplest settings and design a $(12+\varepsilon)$-approximation algorithm, namely \emph{\underline{L}ocal \underline{S}earch \underline{D}eadline \underline{S}cheduling (\lsds)}. At a high level, our technical contribution involves developing an `interval scheduling viewpoint' of the \dpmss problem (see \S~\ref{sec:intervals}). We show that any solution to the \dpmss problem can be thought of as (1) selecting disjoint sub-intervals of a scheduling time-horizon (2) for each selected sub-interal, determining an assignment of packets to the RUs under certain bandwidth restrictions such that all assigned packets start transmission at beginning and finish on or before the end of the sub-interval. We develop a \emph{local search} based algorithm (\lsds) for (1) in conjunction with a \emph{budgeted version of the classical maximum weight bipartite matching} for (2) (see \S~\ref{sec:apx-alg}). Below, we summarize our main contributions.

\label{sec:contribution}
\begin{enumerate}[leftmargin=*]
    \item We utilize WiFi 6 for efficient operation of a smart factory. We model such an operation through a novel variant of classical scheduling problem with release times and deadlines on heterogeneous machines, which we call \dpmss. The scheduler aims to maximize total profit of packets successfully transmitted, where profits indicate relative criticality of packets.
    
    %We assign profits to the applications in a way that more critical the application, higher is the profit. We model the operation of such a smart factory as a scheduling problem that schedules packets from applications based upon the deadline and profit associated with the application. We reduce the WiFi packet transmission problem to a novel variant of deadline scheduling problem \dpmss. 
    
    \item We show that \dpmss is \emph{strongly NP-hard} and use an \emph{admission control} strategy based on \emph{iterative search} to get the \emph{first} approximation algorithm (\lsds) with an approximation ratio $(12+\varepsilon)$, for any given positive $\varepsilon$. In fact, for a special case when the RU configurations for each transmission are given and fixed, our algorithm (namely \lsdsf) is a 12-approximation.
    
    \item We create a simulation framework of factory IoT settings. We reproduce $4$ different use cases under practical channel conditions.     We evaluate the performance of \lsds, \lsdsf and compare with three other benchmarks %namely Earliest Deadline First (EDF), Largest Profit-To-Deadline Ratio First (LRF), Non-Starving variant of Largest Profit-To-Deadline Ratio First (NLRF).  The evaluation is performed  in terms of the ratio between the profit achieved by the algorithms to the total profit of all packets, the percentage of packet drop for both critical and non-critical packets, and runtime. 
    We observe that \lsds outperforms all the benchmarks in terms of profit achieved and packet drops (including critical packets). Further, we observe that \lsds succeeds in determining the schedules in real time. Finally, we observe that along with factory applications \lsds can handle best-effort traffic by utilizing the free resources.

    \end{enumerate}

We have also made the source code available at \cite{dpmss-code} to reproduce all our experiments.

\section{Related Work \& Motivation}
\label{sec:rel}
%In this section, we first present the related work and then discuss the motivation of our work.
\vspace{-0.15cm}

\subsection{Related Work}

%We divide the prior work into two categories (1) WiFi in Industrial settings, (2) optimal sub-carrier allocation.

\noindent \textbf{WiFi in Industrial settings}: WiFi deployment in the industrial space suffers from two major obstacles. Non-determinism associated with the randomised contention based channel access scheme employed by WiFi and unreliability of wireless links.
Multiple works have attempted to overcome the problem of non-deterministic access and unreliable wireless links by modifying the WiFi protocol and/or better scheduling of packets. For example, Seno et al. \cite{seno_enhancing_2017} attempted to enhance determinism in WiFi communications by introducing a coordinator node to enable soft real time applications in industries. %The work \cite{80211n} utilizes space-time block coding (STBC) scheme in Multiple-Input and Multiple-Output (MIMO) transmissions to reduce the unreliability of channel.
%The use of co-operative diversity in a factory environment was investigated in \cite{relaying}, where additional relay nodes were installed to improve reliability of packet delivery. 
The work in \cite{cena_seamless_2016} proposes Wi-Red, a parallel redundancy protocol which aims to improve timeliness and dependability in links by transmitting duplicate packets on redundant links. These works are orthogonal to our approach of utilizing OFDMA for deadline constrained settings. Similarly, a time division multiplexing-based modification of the WiFi protocol, known as RT-WiFi, was proposed in \cite{rt-wifi}. None of these techniques study the utilization of OFDMA for such purposes. Although the work \cite{rt-wifi-ax} performs the job of resource allocation using OFDMA in WiFi 6, it uses a heuristic without providing any performance guarantees.

\noindent \textbf{Optimal sub-carrier allocation: }
Scheduling and resource allocation in IEEE 802.11ax  has gained a lot of attention in recent times~\cite{recursive-infocom18, fairsplit, inamullah_will_2020}, where the authors formulate this as an optimization problem and propose various solutions for the same. Works such as~\cite{fairsplit, recursive-infocom18} propose various algorithms to solve the problem of maximizing the weighted sum rate. %In~\cite{throughput-max}, the authors use an optimization solver for solving their optimization problem. 
%In~\cite{deepmux}, the authors employ  deep neural networks to solve the problem of allocating RU's to users.% In our prior work~\cite{mmru}, we proposed an algorithm namely \texttt{MMRU-ALLOC} that optimally allocates resources to users that minimizes the OFDMA transmission time. Here, we assumed that the set of users to participate in an OFDMA transmission is known. Next, we worked on also selecting the set of users while maximizing the weighted sum rate~\cite{fairsplit}
 %such as the sum of proportional fair rates.  We proposed an approximation algorithm namely \texttt{FairSplit} for the same. 
 Unlike our paper, none of these studies consider the deadlines of each user.
The closest work to ours is by Inam et al. \cite{inamullah_will_2020}, where they designed a scheduler to minimize the number of packet drops in a deadline-driven environment. %Their work first provides an algorithm for estimating the deadlines of packets using Little's Theorem and the Buffer Status Reports collected by the Access Point. Once the deadlines have been estimated, they suggest a simple heuristic that sorts the stations based on their deadlines (earliest first) and then perform an exhaustive search over all possible RU configurations to find the configuration that results in the least number of packets dropped. 
However, lack of theoretical guarantees makes their heuristic less suitable for safety-critical settings.

\noindent
{\bf Scheduling with Deadlines:} Scheduling jobs with deadlines to maximize profit (more commonly called weighted throughput in literature) has been widely studied under a variety of settings. The general setting is consists of a set of jobs, each with a release time, a deadline and a profit. The task is to schedule these jobs on a set of parallel heterogeneous machines to maximize profit of jobs that finish on or before their deadlines. The problem had been established to be strongly NP-hard even in the basic setting of a single machine and unit profit for all jobs~\cite{GareyJ77}. The work of Bar-Noy et al.~\cite{Bar-NoyGNS99} gives the first constant factor approximation algorithms for the problem on multiple machines and arbitrary profits, which was later improved to 2-approximation independently by Bar-Noy et al.~\cite{Bar-Noy01} and Berman et al.~\cite{BermanD00}. For the case with uniform profits, Chuzhoy et al.~\cite{ChuzhoyOR06} improves it further to $1.582$. Throughput maximization on unrelated machines with machine dependent release dates and deadlines has also been studied in the context of wireless sensor networks~\cite{AlayevCHBPL4}.

Note that 5G in cellular networks also uses OFDMA, and deadline scheduling in the context of 5G's ultra-reliable low latency communication (URLLC) solves a similar problem. However, the resource blocks/units can be independently used in 5G~\cite{5g-scheduling, raviv2022joint}, without synchronizing their start times of transmission for each user, unlike in WiFi 6 \cite{11ax-standard}. Requiring such synchronized transmission effectively requires a new class of scheduling algorithms [28]. Unlike 5G, which has access to servers, WiFi’s scheduling happens entirely on access points with small amounts of compute power. This requires algorithms with low computational overhead \cite{deepmux}. Finally, WiFi 6 provides a specific way of splitting the channel into smaller RUs and maximum one RU can be allocated to a node, making it impossible to directly use the algorithms designed for scheduling in URLLC.

To the best of our knowledge, the exact scheduling problem \dpmss has not been studied so far. In particular, two major components in our problem makes the existing algorithms difficult to be used - (1) In \dpmss, the processing of jobs needs to happen in batches in a synchronized fashion on each machine and (2) There is a bandwidth restriction on the machines processing the jobs in a particular batch. We carefully design a novel algorithm that can handle both these constraints effectively with only a constant loss in the approximation factor.

% The exhaustive search uses a recursive expression to estimate the packet drops for a given RU configuration by taking into account the estimated deadlines and the waiting times for the stations. The authors first evaluate the performance of their deadline estimation algorithm against an ORACLE that knows the actual deadlines. The deadline values estimated by their algorithm periodically converge to the actual deadline with a fixed maximum error. Then they compare the performance of their scheduler against the MINUL scheduler which minimizes the upload time by maximizing the throughput. The authors report a $6\times$ reduction in the observed packet drops compared to MINUL.
\subsection{Requirement of a Scheduling Algorithm}

\begin{figure}[!ht]
    \centering
    \includegraphics[width=0.26\textwidth]{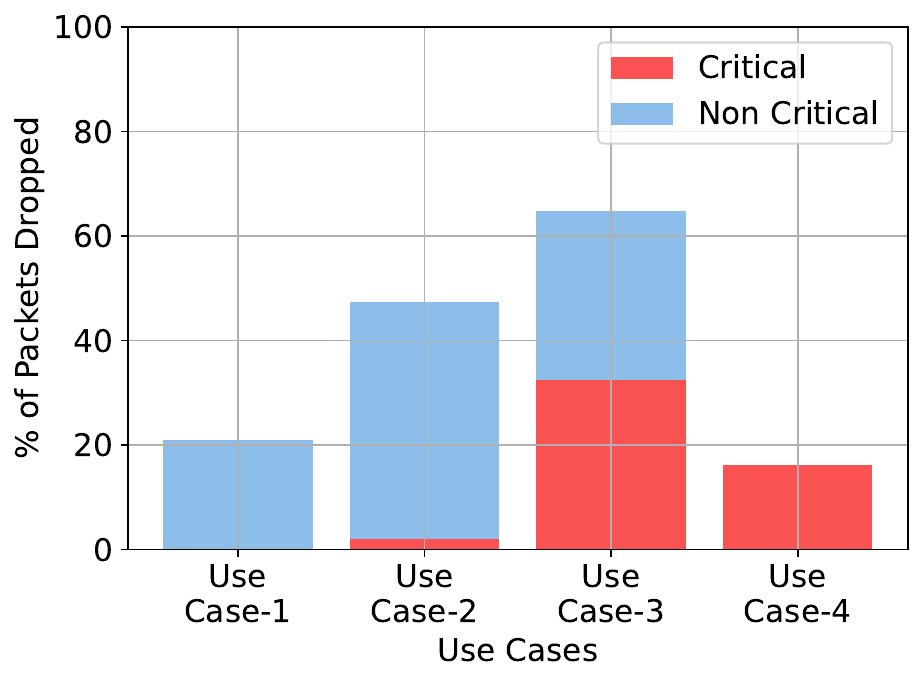}
    \vspace{-0.15in}
    \caption{Percentage of total and critical packets dropped in a total of four different use cases using Earliest Deadline First (EDF). Use Case 1 does not have any critical packets.}
      \vspace{-0.15in}
    \label{fig:edf_failure}
\end{figure}

We now motivate the requirement of new algorithms to schedule WiFi packets with deadline by demonstrating that naive ideas lead to a significant drop in critical packets in an industrial setting. We consider a single WiFi AP and a total of four different use cases, with the total number of devices (apart from the AP) ranging from 40-90, drawing from scenarios mentioned in the research literature (further details in \S5.1). The devices send data to collect control information or monitor the working (additional details in \S4). We use the earliest deadline first (EDF) technique as proposed in prior work~\cite{inamullah_will_2020}, and observe the number of packets dropped (Fig. \ref{fig:edf_failure}). In the last three use cases, the packets from some users are more critical than the other packets. In these cases, we also plot the percentage of the most critical packets dropped.
We observe that the number of packets dropped exceeds $10$\% in all cases. Furthermore, for Use Case-4, we find that all the packets dropped by EDF are critical. This indicates the need for a more intelligent scheduler. \textcolor{black}{Furthermore, a significant number of total packets are dropped in each case. This is because EDF only prioritizes the earliest deadline first without considering how much parallelism can be used. Thus satisfying the deadline of one packet can lead to a cascade of other packets missing their deadlines, as there is a requirement of synchronized transmission.}

Thus, designing an intelligent scheduler is non-trivial, as it needs to take into account the overall latency requirements of all the packets. The scheduler must also run in real-time, i.e, the total execution time should not exceed the time-window available for transmission. 

% We now show that the number of possible ways of allocating the RU's to users is intractable in practice.
% In the above use case, we considered a schedule duration of 50 ms for 20 user devices.
% A single user device can be assigned one among 6 types of RU's.
% Thus, a total of 20 user devices across 50 ms could be assigned a total of $6 \times 20 \times 50 = 6000$ possible configurations. It is infeasible to iterate across all these configurations and choose the optimal one among them in real time.
% \ar{The above needs correction. The numbers are clearly not correct.}

%We now  that the number of possible allocations to each user is.

% Fig.~\ref{fig:sample-industry} shows sample applications of a typical factory setting. 
% \begin{figure}[h!]
%   \centering
%     \includegraphics[width=0.5\textwidth]{figs/industrial-iot.pdf}
%     \caption{Applications in Factory Settings}
%     \label{fig:sample-industry}
% \end{figure}
%\input{problem-statement}
\section{Problem Definition} 

In this section, we define a new variant of classical parallel machine scheduling problem which we call \underline{D}eadline-aware \underline{P}arallel \underline{M}achines \underline{S}cheduling with \underline {S}ynchronized start (\dpmss) and show that our WiFi packet scheduling problem reduces to the above.

\subsection{Problem Model}
\label{subsec:model}
%We model the problem as follows.
We have a single WiFi 6 Access Point that allows parallel OFDMA transmissions.
As mentioned earlier, WiFi 6 allows synchronized parallel transmissions, where multiple users can transmit in parallel but the transmissions must begin together. Furthermore, each such transmission must end within a fixed duration of TXOP = $\delta$.

We also have a fixed set of users, where each user generates packets periodically. 
We denote the set of packets (\emph{jobs}) as $J$. A job $j\in J$ has a specific generation time (\emph{release time}) $r_j$ and a \emph{deadline} $d_j$ before which the packet must be delivered. The relative importance of delivery packet $j$ is modelled by assigning non-negative \emph{profits} $w_j, \forall j\in J$.
%Depending on the type of user, delivery of each packet (completion of each job) may have different level of importance, denoted by $w_j$.
%{\color {blue} @Arani : I do not think the following part in red should come here. This should have already been introduced in the introduction. This section should be more about precise definitions and notations }
%{\color{red} For example, a packet that carries safety-critical information has higher value of $w_j$ than one that carries information about maintenance of equipment.
%We incorporate $w_j$ into our scheduling algorithm by considering it as profit received if the packet is delivered by the deadline.
%Note that delivering a packet after the deadline is equivalent to dropping it as both these cases do not accumulate any profit.}

%We now map the RU's available in WiFi 6.
%As stated earlier, WiFi 6 allows multiple predefined configurations of RU's to be used.
%Each such configuration leads to a different number of parallel transmissions as well as different delivery times of the packets. First, let us consider a simpler setting where a same fixed RU configuration is used for all transmission. 
The RU's are modeled as a set of \emph{machines} $M$ that can \emph{process} jobs $J$. We denote the time taken using  RU (machine) $i \in M$ to deliver (process) a packet (job) $j \in J$ by $p_{ij}$. Further, each machine occupies a \emph{bandwidth} $b_i , \forall i\in M$, during the entire time for which it is active and $B$ denotes the total available bandwidth
%{\color{blue} This is incomplete. The major part of deadline and synchronized start of each batch is totally missing. In fact, that is precisely what makes out scheduling problem new and non-trivial. I am adding back this part back in red. Please check.
\begin{definition} \dpmss :
\label{def:dpmss}
The \dpmss problem requires to schedule jobs non-preemptively in batches such that
\begin{enumerate}
    \item Each batch consists of at most $|M|$ jobs which are to be processed non-preemptively on individual machine (one can think of a batch as one parallel transmission of packets)
    
    \item For each batch, all jobs need to start processing at exactly the same time (hence synchronized start). The endtime of a batch is defined by the latest time for a job in the batch to finish processing
    
    \item The total bandwidth requirement of machines active within a batch must not exceed $B$. 
\end{enumerate}
\end{definition}

The objective is to maximize the total profit of jobs that finish at or before their respective deadlines. One possible approach of solving \dpmss is to greedily group jobs according to deadlines/profits and assign each group to suitable machines. However, such myopic algorithms have arbitrarily bad approximation ratio. In \S~\ref{sec:intervals}, we develop an alternate viewpoint of \dpmss based on scheduling the parallel transmission intervals. This allows us to use a combination of admission control based iterative search.%, leading to provably good algorithms.

\subsection{An Interval Scheduling Viewpoint}
\label{sec:intervals}
We now formalize the interval scheduling viewpoint of \dpmss defined in \S~\ref{subsec:model}. Let $\tT$ be a time-horizon which is pre-defined such that the scheduling needs to happen in the interval $[0, \tT]$ (we assume $\tT$ is an integer). An interval $\I(t_1, t_2)$ is a subset of $[0, \tT]$, where both $t_1, t_2$ are non-negative integers and $t_1 < t_2$. An interval of unit length of the form $[t, t+1], t\in {\mathbb N}$ is called a \emph{time-slot}. We assume that all release dates and deadlines are non-negative integers inside the time-horizon $[0, \tT]$.

\begin{definition} Admissible jobs to a machine:
\label{def:admissible}
A job $j$ is \emph{admissible} to machine $i\in M$ in interval $\I(t_1, t_2)$ if $r_j \leq t_1$ and $t_1 + p_{ij} \leq \min\{t_2, d_j\}$.
\end{definition}
The first condition comes from the fact that the job can only start after it is released. The second condition follows from the fact that if a job is processed by a machine in an interval, then it must complete by the end-time of the interval or it's deadline, whichever is smaller. 

\begin{definition} Admissible jobs to an interval:
\label{def:admissible2}
A job $j$ is admissible to an interval $\I$ if there exists a machine $i\in M$ such that $j$ is admissible to $i$ in interval $\I$.
\end{definition}

Intuitively, one can think of a solution to \dpmss as a collection of intervals within which each batch of parallel transmission happens.

\begin{observation}
\label{obs:dpmss-interval}
Any feasible solution to \dpmss consists of %a set of disjoint intervals $\I_1, \I_2, \cdots \I_k$ and sets of admissible jobs $J(\I_r)\subseteq J, \forall r=1,2,\cdots k$.

%The above observation allows us to look at the \pmss problem as
\begin{enumerate}
\item Disjoint intervals of $[0, \tT]$, $\I_1, \I_2, \cdots \I_k$, and
\item for each interval $I_r, r=1,2,\cdots k$, a set of admissible jobs $J(\I_r)$ and a set of active machines whose total bandwidth does not exceed $B$, along with an assignment of the jobs to active machines inside $\I_r$.
\end{enumerate}
such that for any interval $\I_r$, no machine is assigned more than one job in the set $J(\I_r)$ and 
no job belongs to both $J(\I_r), J(\I_{r'}), r\neq r'$.
\end{observation}

Hence, we need to design an algorithm that determines (1) and (2) such that the total profit of jobs scheduled is maximized. We shall adopt this point-of-view of \dpmss in the following discussion.

\begin{theorem}
\label{thm:hardness}
The \dpmss problem is strongly NP-hard
\end{theorem}
\begin{proof}
Let us consider the simplest setting of \dpmss where we have only machine of unit bandwidth and the total bandwidth available $B=1$. Then the \dpmss problem reduces to the classical decision problem  of single machine non-preemptive scheduling with release dates and deadlines, where task is to decide whether all jobs can be scheduled within their respective deadlines or not. This problem has been proved to be strongly NP-hard in~\cite{GareyJ77}. Hence \dpmss is strongly NP-hard.
\end{proof}

We next establish that even a sub-problem to our main problem is computationally hard.
\begin{theorem}
\label{thm:budgeted-matching}
For a fixed interval $\I$ and a given bandwidth $B$, determining the maximum profit subset of admissible jobs and their assignment to machines such that total bandwidth of machines involved in the assignment does not exceed $B$ is weakly NP-hard.
\end{theorem}
The above theorem follows via a simple reduction of the Knapsack problem to this problem. On the positive side, the above problem admits a  PTAS~\cite{BergerBGS11}.

\section{Algorithm for \dpmss}
\label{sec:apx-alg}
Given the above hardness results, we give approximation algorithms for \dpmss. We first consider a slightly easier setting where the set of machines (RUs) to be used for processing is the same fixed one for each batch - we call this \dpmss with \emph{fixed} configuration (\dpmssf). In other words, we assume that we are given a valid configuration of machines that satisfies constraint (3) in Definition~\ref{def:dpmss}. Note that Theorem~\ref{thm:hardness} directly implies that even this simpler setting is strongly NP-hard. In \S.~\ref{sec:extension}, we show how to extend this algorithm to the general setting where the algorithm must decide which machines to engage for each batch under bandwidth constraint.

\subsection{Algorithms for \dpmssf}
We need a few more notations in order to describe our algorithm. For any subset of jobs $\JA\subseteq J$, let $w(\JA) = \sum_{j\in \JA} w(j)$ denote the total profit of all jobs in $\JA$. Let $\delta$ be the maximum length of any interval in any feasible solution. In general, $1\leq \delta \leq \tT$, for our applications, $\delta$ (duration of one transmission) is a small constant. 

%\begin{definition}
%Given a partial schedule and an interval $\I$ along with a set of jobs $J(\I)$ that are admissible in $\I$, a scheduled job $j$ conflicts with $J(\I)$ if  
%\end{definition}

\begin{definition} Conflicts :
An interval $\I_1=[t_1, t_2]$ is defined to be in conflict with an interval $\I_2 = [t_3, t_4 ]$ if either $t_1 \leq t_3 \leq t_2$ or $t_3 \leq t_1 \leq t_4$.   
\end{definition}

\textcolor{black}{We show our main algorithm named as \lsdsf (Local Search Deadline Scheduling Fixed split) in  Algorithm~\ref{alg:pmss}}. %\ar{Let's specify line numbers of Algorithm 1 in this para.} 
The main idea is the following. We start with an empty set of intervals $\SA$ and empty set of jobs $\JA$ - both of which would eventually form our solution. The outer loop of Line 3 iterates over all possible integral lengths of the intervals 1 to $\delta$. For a fixed length $\ell$, Line 4 of the algorithm iterates over all possible intervals of length $\ell$. While considering a particular interval, say $\I=[t, t+\ell]$, the algorithm first determines the admissible set of jobs in $\I$ with \emph{maximum possible profit} that have not yet been added to $\JA$ - let $J(\I)$ denote this set of jobs. Solving this sub-problem itself is non-trivial since there are exponentially many subsets to be considered. However, it reduces to an instance of the classical \emph{maximum weight bipartite matching problem}. We invoke standard algorithms to determine the maximum profit subset of jobs $J(\I)$ and also the assignment of each job $j\in J(\I)$ to individual machines $i\in M$ such that $j$ is admissible to $i$ in interval $\I$ (Line 5) . This sub-procedure is listed in Algorithm~\ref{alg:maxprofit}. Lines 6-11 form the heart of our algorithm. We need to decide whether to add the new interval $\I$ to $\calS$. To this end, we check whether $w(J(\I))$ is strictly bigger than twice the total profit of already selected jobs in intervals that are in conflict with $\I$ and are currently in $\calS$ (Line 8). If this is true, then we discard all the conflicting intervals from $\calS$, remove all jobs assigned to such intervals from $\JA$ while adding new interval $\I$ to $\calS$ and a new set of jobs $J(\I)$ (Lines 9-10). Otherwise, we ignore the interval $\I$.

\begin{algorithm}[t]
\small
    \caption{Algorithm \lsdsf $(J, M, \tT, \delta)$}
    \label{alg:pmss}
    \begin{algorithmic}[1]
        % \STATE {\bfseries \ALG$(\calI)$} 
        %\STATE Run a $\rho$-approximation algorithm for \vkp on $\cI$ --- let $\cC$ be the set of centers.
        \STATE $\SA = \{\}, \JA = \{\}$

        \FOR {$\ell = 1,2,\cdots \delta$} 
            \FOR {$t=0,1,2,\cdots \tT-\ell$}
                \STATE Let $\I$ : Interval $[t, t+\ell]$ 
                \STATE $(\calM, J(\I)) \leftarrow$ \maxpft$(J\setminus \JA, \I, M)$
                \STATE Let $\calS' = \{\I' \in \SA : \I' \text{ conflicts with } \I \} $
                \STATE Let $J' = \bigcup_{\I' \in \calS'} J(\I')$
                \IF {$w(J(\I)) > 2w(J')$}
                    \STATE $\JA \leftarrow \JA\setminus J' \cup J(\I)$
                    \STATE $\SA \leftarrow \SA \setminus \calS' \cup \{\I\}$
                    \STATE Assign job $j$ to machine $i$ inside interval $\I$ if and only if $(ji)$ is in matching $\calM$  
                \ENDIF
            \ENDFOR
        \ENDFOR
    \end{algorithmic}
\end{algorithm}

\begin{algorithm}[t]
\small
    \caption{Algorithm for \maxpft $(J', \I, M)$}
    \label{alg:maxprofit}
    \begin{algorithmic}[1]
        % \STATE {\bfseries \ALG$(\calI)$} 
        %\STATE Run a $\rho$-approximation algorithm for \vkp on $\cI$ --- let $\cC$ be the set of centers.
        %\STATE $\calS = \{\}, \JA = \{\}$
        %\STATE $\JA' = \{j\in \JA : j \text{ is admissible to \I }\}$
        \STATE $J(\I, i) = \{j\in J' : j \text{ is admissible to } i \text{ in } \I \} $
        \STATE $J' = \bigcup_{i\in M} J(\I, i) $
        \STATE Construct bipartite graph $G$ where \\
        (a) $J'$ and $M$ are the two partite sets of vertices \\
        (b) For $j\in J'$ and $i\in M$ add edge $(ji)$ with weight $w_{j}$ if and only if $j\in J(\I,i)$
        
        \STATE Solve Maximum Weight Bipartite Matching on $G$ - let $\calM$ be the matching and $\tilde{J}$ be the set of jobs matched 
        
        \STATE Return $\calM, \tilde{J}$
    \end{algorithmic}
\end{algorithm}

%\ar{For any given interval $\I$, the maximum weighted matching algorithm can choose the optimal subset of jobs that should be scheduled in it.}

%\ar{The fact that if intervals are pre-decided, the maximum weighted matching algorithm can choose the optimal jobs  is not clear from any prior discussion. I think we need to explain this somewhere.}

\subsection{Analysis}
\noindent
{\bf Runtime.} \textcolor{black}{\lsdsf} showed in Algorithm~\ref{alg:pmss} runs in time $\bigO(\tT\cdot\delta\cdot \rho(|J|, |M|)$ where, $\rho(|J|, |M|)$ is the runtime of any algorithm for maximum weight bipartite matching. The currently best known theoretical bound for $\rho$ is nearly linear in $|J|\cdot |M|$~\cite{ChenKLPGS22}. However, one can use a more practical algorithm running in similar time with only an $(1+\varepsilon)$-factor loss in the approximation ratio (think of $\varepsilon$ as a tiny positive quantity)~\cite{Duan14}.

\noindent
{\bf Approximation Ratio.} We are going to prove the following theorem in this section. We defer the proof of some of the lemmas to an extended version of the paper due to lack of space.

%\ar{Why mix time complexity and approximation ratio here? I would say that let the theorem talk only about approximation ratio.}
\begin{theorem}
\label{thm:main}
\lsdsf is a 12-approximation for the \dpmssf problem.
\end{theorem}

%Here, $\rho(|J|, |M|)$ is the runtime of any algorithm for maximum weight bipartite matching. The currently best known theoretical bound for $\rho$ is nearly linear in $|J|\cdot |M|$~\cite{}. However, one can use a more practical algorithm running in similar time with only an $(1+\varepsilon)$-factor loss in the approximation ratio (think of $\varepsilon$ as a tiny positive quantity)~\cite{}.
%\ar{Here, the exact constant factor might be important?}

%\noindent
%{\bf High Level Overview.} 
%The following lemma follows in a straightforward way from the algorithm description and Observation~\ref{obs:dpmss-interval}. We skip a formal proof due to lack of space. \ar{Saying this is risky.}

\begin{lemma}
\label{lem:feasibility}
\textcolor{black}{\lsdsf} always computes a feasible solution to a given \dpmssf instance.
\end{lemma}

\begin{proof}
Lines 6-10 ensure that the algorithm never selects two conflicting intervals. Furthermore, \maxpft ensures that for any interval $\I=[t_1, t_2]$ added to $\SA$, only admissible jobs are selected in $J(\I)$, which implies that any such job $j$ can start processing at $t_1$ and finish on or before $\min\{t_2, d_j\}$.
\end{proof}

% of all, it is not difficult to show that our algrorithm always produces a feasible solution to \dpmss. 

The more non-trivial part is to prove the approximation ratio.
First, we prove a lemma about the \maxpft subroutine.

\begin{lemma}
\label{lem:maxpft}
Given an interval $\I$, a subset of jobs $J'$ and a set of machines $M$, the algorithm \maxpft returns the subset of admissible jobs of $J'$ in $\I$ with maximum possible profit
\end{lemma}

\begin{proof}
By definition of $J'$ in Lines 2-3 of \maxpft, it is ensured that the jobs considered as possible candidates to be scheduled in interval $\I$ are all admissible. Furthermore, while constructing the bipartite graph $G$, we introduce an edge $(ji)$ for $j\in J', i\in M$ if and only if $j$ is admissible to $i$ inside $\I$. Hence it follows that a maximum weight bipartite matching on this graph gives the maximum profit admissible set of jobs in interval $\I$.
\end{proof}

Let $\sS$ denote the set of intervals $\Is_1, \Is_2, \cdots \Is_{k'}$ picked by some \emph{optimal} solution to the given \dpmss instance along with corresponding sets of admissible jobs $\bJ(\Is_r), r=1,2,\cdots k'$ and their assignment to machines within each interval. Also 
Let $\Js$ be the set of jobs in an optimal solution to \dpmss.
For the purpose of analysis, we divide the set $\Js$ in to three disjoint sets according to \textcolor{black}{\lsdsf} Algorithm~\ref{alg:pmss}. Recall that $\JA$ denotes the set of jobs at the end of \lsdsf. Let $\Js_1$ denotes the set of jobs that were never added to set $\JA$, $\Js_2$ denotes the set of jobs that were added to set $\JA$ at some iteration (Line 9-10), but discarded at a later iteration due to conflict with a new interal (Lines 7-8) and $\Js_3$ is the remaining set of jobs that are part of both $\JA$ and $\Js$. We are going to show that $w(\Js_1) \leq 2w(\JA)$ and $w(\Js_2) \geq 2w(\JA)$. 
Finally, let $\tJ$ denote the set of jobs that are in $\JA$ but not in $\Js$.
Note that the both the optimal and our algorithm gains the profit of jobs in $\Js_3$ whereas the algorithm does not have the profits of either $\Js_1$ or $\Js_2$. The heart of the analysis is to prove that the total profit gained by jobs in $\JA$ scheduled by the algorithm, is at least a constant fraction of the profit gained by each of the latter two sets in optimal. But first we show that the total profit of jobs in $\Js_1$ is only a constant times larger than the total profit of jobs in $\Js_2, \Js_3$ and $\tJ$.
We first prove the following lemma about $\Js_2$.

\begin{lemma}
\label{lem:OPT-1}
$w(\Js_1) \leq 5((w(\Js_2) + w(\Js_3) + w(\tJ))$
\end{lemma}

\begin{proof}
We partition set $\Js_1$ in to disjoint sets $\Js_{1,1}, \cdots \Js_{1,q}$ such that they are scheduled in intervals $\tI_1, \tI_2, \cdots \tI_q$ respectively in the optimal solution. Now consider a particular interval $\tI_i$ for any $i=1,2,\cdots q$. Lines 2-3 of the algorithm iterate over all possible intervals of length at most $\delta$ and hence the specific interval $\tI_i$ much have been considered at some iteration of the algorithm. Now two cases could happen as follows.
\begin{itemize}[leftmargin=*]

    \item Case 1 : Suppose the interval $\tI_i$ was actually added to $\SA$ along with a subset of admissible and currently unscheduled jobs $J(\tI_i)$. %admissible to $\tI_i$ returned by the subroutine \maxpft. 
    Crucially, $\Js_{1,i}$ is a candidate set for \maxpft since these jobs are not part of the algorithm's set by definition of $\Js_1$. Since \maxpft returns the set with maximum profit, $ w(\Js_{1,1} \leq w(J(\tI_i) $.
    
    %Also, by definition, $J(\tI_i)$ belongs to one of the sets $\Js_2, \Js_3$ or $\tJ$. 
    
    \item Case 2 : Suppose the interval $\tI_i$ was not added to $\SA$. Let $\tS$ denote the set of intervals currently in $S$ that are in conflict with $\tI_i$ (as per Line 6) and let $J' = \bigcup_{I'\in \tS} J(I')$ (Line 7). Hence, by condition of Line 8, $w(\Js_{1,i}) \geq 2w(J')$. 
\end{itemize}

Now we sum up the inequalities for all sets $\Js_{1,1}, \Js_{1,2}, \cdots \Js_{1,q}$. We make the following observations. Consider any interval $\hI$ ever added to the algorithm and its corresponding jobs $J(\hI)$. Then,
\begin{enumerate}[leftmargin=*]
    \item $J(\hI)$ appears in the RHS for exactly one inequality for Case 1
    
    \item $J(\hI)$ appears in the RHS for at most two inequalities for Case 2. The reason for this is the following. The intervals $\tI_1, \tI_2, \cdots \tI_q$ are disjoint since they form part of an optimal solution. Further, by design of our algorithm, if the interval $\hI$ belongs to the conflict set for a specific interval $\tI_i$, then the length of $\hI$ is at most the length of $\tI_i$. The above two facts imply $\hI$ can conflict with at most two intervals among  $\tI_1, \tI_2, \cdots \tI_q$ and we are done.
    \end{enumerate}
    Finally note that every job ever added to $\JA$ (and possibly discarded later) belong to one of the sets $\Js_2, \Js_3, \tJ$.
    Hence it follows that 
    \[ w(\Js_1) = \sum_{i=1}^q \Js_{1,i} \leq 5(w (\Js_2) + w(\Js_3) + w(\tJ) \]
    
\end{proof}

We now prove that the total profit of jobs in $\Js_2$ is at most the total profit gained by the jobs in $\JA$. 
%\begin{lemma}
%\label{lem:OPT-2}
%$w(\Js_2) \leq w(\JA)$
%\end{lemma}
We do so by proving a more general statement that will imply the above lemma as a corollary.

\begin{lemma}
\label{lem:discarded}
Let $\tS$ be the set of all intervals that were added but later discarded by Algorithm 1. Then  
$w(\JA) \geq w(\bigcup_{\I\in \tS} J(\I))$
\end{lemma}
The proof of the above lemma requires a delicate `charging' argument which we are going to develop over a few steps. We first define a collection of tree structures that models the behavior of the algorithm and allows us to bound the profit of intervals we discard.

\begin{definition} Charging Forest: We define a node for each interval in $\SA$ and $\tS$. We make an interval node $\I$ the parent of another node $\I'$ if and only if $\I'$ was discarded by the algorithm from $\JA$ due to conflict with $\I$.  
\end{definition}

\begin{observation}
\label{obs:roots}
The roots of the charging forest are precisely the intervals in $\JA$ upon termination, while the internal nodes and the leaves form the set $\SA$.
\end{observation}

The following lemma forms the heart of our analysis,
\begin{lemma}
\label{lem:geomsum}
For any internal node of the charging forest, $\I$, the total profit of the jobs corresponding to the sub-tree rooted at $\I$ excluding the node $I$ is at most the profit of $\I$.
\end{lemma}

\begin{proof}
We define levels of nodes in the charging forest $\calF$ in the natural way. The leaf nodes are level 0. An internal node is of level $j+1$, where $j$ is the maximum level of any of it's children. We prove the lemma by induction on the level. The hypothesis is clearly true for level 0. Now let it be true for all level $0,1,2,\cdots j$ and consider a node $\I$ at level $j+1$. Let $\I'_1, \I'_2 \cdots \I'_p$ be it's children. By construction of the charging forest and Line 8 of the algorithm, 
\[ w(J(\I)) > 2\sum_{i=1}^p w(J(\I'_i)) \]

Further, applying induction hypothesis,  the total profit of jobs corresponding to the sub-trees rooted at the nodes  $\I'_1, \I'_2, \cdots \I'_p$ excluding the nodes themselves is at most $\sum_{i=1}^p w(J(\I'_i)$. Hence, adding the total profit of all jobs in the subtree rooted at $\I$ excluding itself is at most $2\cdot w(J(\I))/2 = w(J(\I))$
\end{proof}

Hence, using Lemma~\ref{lem:geomsum} on the roots of the charging forest and Observation~\ref{obs:roots}, we have Lemma~\ref{lem:discarded}. As a corollary of Lemma~\ref{lem:discarded}, we get the following lemma:

\begin{lemma}
\label{lem:OPT-2}
$w(\Js_2) \leq w(\JA)$
\end{lemma}
%\begin{proof}
%The only observation we require is that $\Js_2$ is a subset of jobs in $\bigcup_{\I\in \tS} J(\I)$. 
%\end{proof}

Now we finish the proof of Theorem~\ref{thm:main}. Using Lemmas~\ref{lem:OPT-1}, ~\ref{lem:OPT-2}, 
\begin{align*}
w(\Js) &= w(\Js_1) + w(\Js_2) + w(\Js_3) \\
&\leq 5(w(\Js_2) + w(\Js_3) + w(\tJ))+ w(\Js_2) + w(\Js_3) \\
&\leq 10w(\JA) + 2(\JA) = 12w(\JA)
\end{align*}

\subsection{Extension to DPMSS.}
\label{sec:extension}
%In the above exposition, we have assumed that the set of machines is fixed for every interval. However, recall that in our definition of \dpmss, each machine $i\in M$ occupies a certain bandwidth $b_i$ and we are given a total bandwidth of $B$. The scheduler now has the additional task of selecting a suitable set of machines for each batch without exceeding a total bandwidth of $B$. 
We briefly discuss how to extend \textcolor{black}{\lsdsf} to \dpmss. The only difference lies in the sub-routine \maxpft. In case of \textcolor{black}{\dpmssf}, this subproblem could be solved using maximum weight bipartite matching over the set of machines and admissible jobs. In case of \dpmss, we need to solve a budgeted version of this problem since the scheduler has the additional task of allocating the total bandwidth $B$ among machines. Formally, we again construct the bipartite graph $G$ exactly as in Algorithm~\ref{alg:maxprofit}. However, we need to ensure that the total bandwidth of nodes selected by the matching $\calM$ from the set $M$ does not exceed $B$. As stated in Theorem~\ref{thm:budgeted-matching}, this problem itself is weakly NP-hard (unlike the unbudgeted case) and there is PTAS~\cite{BergerBGS11} known for this problem. %The theoretically best known approximation ratio is $(1+\varepsilon)$ for any fixed positive $\varepsilon$. 
Plugging in the latter gives us the following theorem in a straightforward fashion.

\begin{theorem}
\label{thm:bdpmss}
There is a $(12+\varepsilon)$-approximation algorithm (\textcolor{black}{\lsds)} for \dpmss, for any fixed $\varepsilon > 0$.
\end{theorem}

\noindent
{\bf Runtime.} The runtime of \textcolor{black}{\lsds} with the above modification is $\bigO(\tT\cdot\delta\cdot \rho'(|J|, |M|)$
Here $\rho'$ is the runtime of the budgeted maximum weighted bipartite matching algorithm used. Using the state-of-the-art~\cite{BergerBGS11}, this can be implemented in time  $(|J|\cdot |M|)^{\bigO(1/\varepsilon)}$.
%ar{Lets add one paragraph discussion about time complexity of our technique.}
\section{Evaluation Setup}
\label{sec:evaluation}

In this section, we describe our evaluation setup, evaluation test cases, metrics, and the benchmarks. 
%We first describe our simulation setup, then present the baseline algorithms, and finally, we present the evaluation results. 

\noindent \textbf{Simulation Paramters:}
\begin{table}[t]
\centering
\small
%\vspace{-0.15in}
\caption{Simulation Parameters}
\vspace{-0.15in}
\begin{tabular} { 
  | p{2.2cm} | p{1.9cm} || p{2.1cm} | p{0.85cm} | }
 \hline
  \textbf{Parameter} &   \textbf{Value} &   \textbf{Parameter} &   \textbf{Value} \\
 \hline
  Stations  &   use-case specific  &  MCS  &   $11$ \\
 \hline
   Transport Protocol  &   UDP & Bandwidth &   $40$ MHz \\
 \hline
      Frame Aggregation & Disabled &  Guard Interval  &   $3200$ ns  \\
 \hline
 
\end{tabular}
%\vspace{0.05cm}
\vspace{-0.15in}
\label{simulation-setup}
\end{table}
%We first describe our simulation scenario and then present various use cases that we have replayed.
The parameters for our simulation are provided in Table~\ref{simulation-setup}. Note that we simulated for $200$ ms because the periodicity of transmissions allows the WiFi access point to reuse the same schedule repeatedly. While this simulation horizon ensures packets cannot be scheduled across rounds of $200$ ms, a deadline above $200$ ms is considered easily achievable over WiFi and thus this does not hurt our performance. We primarily focus on settings where deadlines are $\leq200$ ms, as these are most challenging for WiFi to handle. Note that this also implies that in case of any changes in the number of nodes, a new schedule can be generated and utilized with a delay of $200$ ms. This delay is considered relatively small, since WiFi nodes often take up to $1$ s \cite{wifi-delay} to establish a connection.

Additionally, our simulation disables frame aggregation since there is no packet queuing at the stations (a new packet arrives only after the deadline for the last packet has expired, implying that it was either transmitted or dropped). We evaluate under both good channel conditions and poor channel conditions. We also assume that the trigger frame, carrying the station-to-RU mapping, along with the Uplink OFDMA data frame is able to finish transmission within TXOP.

\noindent\textbf{Implementation:} We have implemented the code using C++, and compile it using gcc compiler with the optimization flag ``O2". In \S.\ref{sec:extension}, we had implemented the \maxpft subroutine using an existing PTAS for maximum weight budgeted matching~\cite{BergerBGS11} which leads to the desired approximation ratio. However, PTAS-es are prohibitive in practice. In the actual implementation, we use a much faster iterative heuristic which cleverly navigates through the search space of all possible RU configurations. 
%We sketch the main ideas here and defer the details to a full version of the paper. 
The algorithm starts with an infeasible configuration containing the maximum number of RU's of each type (essentially the set of all machines $M$), ignoring the bandwidth restriction and greedily determining the best possible assignment of $M$ jobs to the machines. This step prunes the possible set of jobs to size $|M|$. The next idea is to iterate over all possible configurations and only these pruned set of jobs and output the best combination. The algorithm can be implemented in time $\bigO(\mu\cdot |M|\log |M|)$, where $\mu$ is the total number of possible feasible RU configurations. Since in practice, both $M$ and $\mu$ are quite small (at most $36$ for our use cases), it makes our algorithm much faster in practice compared to the theoretical bounds. %We choose the amount of available bandwidth to be sufficient to ensure that less than $5\%$ of critical packets and $20\%$ of total packets are dropped.
Except for the cases where we separately mention, this is typically equal to $40$MHz.

%For the implementation of heterogeneous RU configurations, we used an iteration over all possible configurations to get the exact solution to DPMSS. This lets us avoid choosing a proper value of $\epsilon$ and gives us the exact solution. As described later, we are able to get faster than real-time performance on a server on all but the most challenging use case. 

\noindent\textbf{Evaluation Metrics:}
The metrics of interest for us are (1) profit ratio which is defined as the ratio between the profits obtained vs total profit that could be achieved if all the packets got scheduled in due time, (2) drop percentage which is defined as the percentage of total packets dropped, and (3) critical drop percentage which is defined as the percentage of critical packets dropped. For each use case, the criticality is decided by the application with the highest profit, and (4) runtime which is defined as the time needed to execute the algorithm.

\subsection{Factory Use Cases}
We now describe various simulated use cases, each of which represents different real-world scenarios. %Each use case has some different properties. For example, Use Case-1 has a high packet generation rate and stringent deadline. Use Case-2 has lower packet generation and looser deadlines. Use Case-3 has a mix of generation rates, but one specific application has a very high profit. Use Case-4 contains a very high generation rate but relatively loose deadlines. Use Case-5 contains a large variety of applications. 
%Together, these use cases represent a large variety of real use cases that are likely to be seen in practice. 
Since some of these use cases generate packets with sizes following certain distributions, we repeat the simulation $100$ times and plot the $95\%$ confidence interval. We assign the profit of each application based on its criticality, i.e. applications with higher reliability requirements get higher profits, as in \cite{raviv2022joint}.

\noindent
\textbf{Use Case-1 (UC-1):} We consider a factory scenario with a set of sensors and controller~\cite{usecase1-paper}. The network simulates real-time communication between the sensors and the controller. %We consider a tree topology with a single controller. %Every sensor communicates with this central controller. 
The communication between the sensor and the controller is based on the application profiles presented in Table.~\ref{usecase1-table}. Each profile includes a periodic transmission rate, a minimum, and maximum frame size, and an end-to-end latency requirement for frames. Each sensor's communication pattern is uniformly distributed from one of the application profiles. We create a  topology with $50$ sensors and $1$ controller as AP. %All the sensors talk to the controller. 
For each device, we set the deadline as the maximum latency corresponding to its profile. We retain the same profit for all nodes, as no specific values are reported.
%\ar{Should we replace the term profit in the table by profits?}

\begin{table}[bt]
    \centering
    \small
    % \vspace{-0.15in}    
     \caption{Use Case-1 (UC-1) Application Characteristics~\cite{usecase1-paper}}
    \vspace{-0.15in}    
    \begin{tabular}{|p{1.05cm}|p{1.34cm}|p{1.6cm}|p{1.1cm}|p{0.7cm}|p{0.8cm}|}
    \hline
        \textbf{Appl.} & \textbf{Gen. rate (pkts/s)}  & \textbf{Size (B)} & \textbf{Deadline (ms)} & \textbf{Profit} & \textbf{\#Nodes} \\ \hline
        Profile 1 & $4000$ & $U(64,$ $128)$ & $0.25$ & $10$ & $10$ \\ \hline
        Profile 2 & $2000$ & $U(128,$ $256)$ & $0.5$ & $10$ & $10$ \\ \hline
        Profile 3 & $1000$ & $U(256,$ $512)$ & $1$ & $10$ & $10$ \\ \hline
        Profile 4 & $500$ & $U(512,$ $1024)$ & $2$ & $10$ & $10$ \\ \hline
        Profile 5 & $250$ & $U(1024, 1522)$ & $4$ & $10$ & $10$ \\ \hline
    \end{tabular}
    
    \vspace{-0.15in}    

    \label{usecase1-table}
\end{table}

\noindent
\textbf{Use Case-2 (UC-2):}
Here, we consider an industrial IoT system with different application settings~\cite{usecase3-paper}. Table.~\ref{usecase3-table} shows the application characteristics. The control traffic has the maximum profit.
\begin{table}[t]
\small
    \centering
     \caption{Use Case-2 (UC-2) Application Characteristics~\cite{usecase3-paper}}
     \vspace{-0.15in}
    \begin{tabular}{|p{1.7cm}|p{1.3cm}|p{0.5cm}|p{1.1cm}|p{0.75cm}|p{1cm}|}
    \hline
        \textbf{Appl.} & \textbf{Gen. rate (pkts/sec)}  & \textbf{Size (B)} & \textbf{Dead-line (ms)} & \textbf{Profit} & \textbf{\# Nodes} \\ \hline
        Smart meters & $1.25$ & $100$ & $16$ & $10$ & $15$ \\ \hline
        Status info & $2.5$ & $100$ & $16$ & $20$ & $15$ \\ \hline
        Reporting \& logging & $0.75$ & $500$ & $1000$ & $30$ & $15$ \\ \hline
        Data polling & $1$ & $500$ & $16$ & $10$ & $15$ \\ \hline
        Control traffic & $937.5$ & $100$ & $16$ & $160$ & $20$ \\ \hline
        Video surveillance & $2000$ & $1500$ & $1000$ & $10$ & $10$ \\ \hline
    \end{tabular}
    %\vspace{0.08cm}
    \label{usecase3-table}
        %\vspace{-0.1in}
\end{table}

\noindent
\textbf{Use Case-3 (UC-3):}
As in ~\cite{usecase4-paper}, we consider another industrial IoT (IIoT) network. Table.~\ref{usecase4-table} represents various applications in this network and their characteristics. Here, the arrival of packets is modeled as a Poisson process with an average arrival rate $\lambda$ packets/sec as mentioned in $2^\text{nd}$ column of Table.~\ref{usecase4-table}. We assign motion-control and robotic-control applications the highest profit followed by collaborative AGV (Automated Guided Vehicle). The asset/process monitoring has the least profit.
\begin{table}[t]
\small
    \centering
     \vspace{-0.15in} 
      \caption{Use Case-3 (UC-3) Application Characteristics~\cite{usecase4-paper}}
    \vspace{-0.15in} 
    
    \begin{tabular}{|p{2cm}|p{1.3cm}|p{0.5cm}|p{1.1cm}|p{0.8cm}|p{0.8cm}|}
    \hline
        \textbf{Appl.} & \textbf{Gen. rate (pkts/s)} & \textbf{Size (B)} & \textbf{Deadline (ms)} & \textbf{Profit} & \textbf{\# Nodes} \\ \hline
        Motion control & $40000$ & $50$ & $1$ & $30$ & $10$ \\ \hline
        Collaborative AGV & $40000$ & $50$ & $4$ & $20$ & $10$ \\ \hline
        Robotic control & $40000$ & $50$ & $10$ & $30$ & $10$ \\ \hline
        Asset/process monitoring & $40000$ & $50$ & $100$ & $5$ & $10$ \\ \hline
    \end{tabular}
    %\vspace{-0.2in}    
    \label{usecase4-table}
    \vspace{-0.15in}
\end{table}

\noindent
\textbf{Use Case-4 (UC-4):}
As in ~\cite{flexible_iot_ieee_2020}, we now consider a metal processing site. It has a set of machines and tools managed in a certain area. %Workers are at positions where they can operate the machines. For operation monitoring and preventive maintenance, sensors are attached to the machines. For operation monitoring, monitor cameras and sensors are installed on machines to monitor the operation status. In the case of preventive maintenance, various sensors are installed on machine tools. In addition, there is a communication requirement for monitoring objects and monitoring the movement of people.  
Table~\ref{usecase5-table} represents the set of applications for the metal processing site, with a total of $10$ applications running on $59$ stations.
%and packet drop penalties have been assigned in the order Human Safety > Equipment Control > Quality Supervision > Factory Resource Management. 
\begin{table}[!ht]
\small
\caption{Use Case-4 (UC-4) Application Characteristics~\cite{flexible_iot_ieee_2020}}
    \vspace{-0.15in}
    \centering
\begin{tabular}{|p{2.1cm}|p{1.27cm}|p{0.7cm}|p{1.1cm}|p{0.7cm}|p{0.8cm}|}
    \hline
      \textbf{Appl.} & \textbf{Gen. rate (pkts/s)} & \textbf{Size (B)} & \textbf{Deadline (ms)} & \textbf{Profit} & \textbf{\# Nodes} \\
     \hline
        Size inspection by line camera (line sensor) & $1$ & $30000$ & $5000$ & $30$ & $3$ \\
     \hline
      Detect defect state & $10$ & $500$ & $500$ & $50$ & $4$ \\
     \hline
      Sensing for managing AC & $0.016$ & $64$ & $6000$ & $45$ & $1$ \\
     \hline
      Preventive
maintenance & $0.02$ & $30$ & $1000$ & $50$ & $2$ \\
     \hline
      Monitoring of equipment & $1$ & $20$ & $1000$ & $30$ & $2$ \\
     \hline
      Counting number of wrench operations & $0.016$ & $64$ & $100$ & $40$ & $10$ \\    
     \hline
     Movement analysis wireless beacon & $2$ & $4000$ & $4000$ & $10$ & $6$ \\    
     \hline
     Racking assets (beacon transmission) information of equipment and things & $1$ & $200$ & $1000$ & $15$ & $20$ \\    
     \hline
     Tracking 
parts, stock  RFID tag & $0.028$ & $1000$  & $1000$ & $45$ & $10$ \\    
     \hline
     Techniques, knowhow from experts video, torque waveforms & $50$ & $24000$ & $200000$ & $1$ & $1$ \\    
     \hline
    \end{tabular} 
    \vspace{-0.15in}
    %\vspace{0.09cm}
    %\vspace{-0.4in}
    \label{usecase5-table}
\end{table}

\noindent
\textbf{Benchmarks.} We consider three benchmarks namely, Earliest Deadline First (EDF), Largest Profit-To-Deadline Ratio First (LRF), and a Non-Starving variant of Largest Profit-To-Deadline Ratio First (NLRF). All three benchmarks sort the stations based on a scheduling metric and assign the first $M$ stations to $M$ RUs. It exhaustively searches over all RU configurations and picks the one that provides the maximum profit.
The scheduling metric ($sm$) used for EDF and LRF are the deadline ($d_j$) and profit-to-deadline ratio ($p_j/d_j$) respectively, whereas for NLRF it is given by:

% is defined as follows:
% For EDF, 
% \begin{equation}
%     sm= Deadline
% \end{equation}
% For LRF, 
% \begin{equation}
%     sm=\frac{Profit}{Deadline}
% \end{equation}
%The NLRF, 
\begin{equation}
    sm_{nlrf}=\frac{p_j / d_j}{(T_{packets} + 1) / (G_{packets} +1)},
\end{equation}
where $T_{packets}$ and $G_{packets}$ represent the number of transmitted and generated packets respectively. This is to ensure that the profit-to-deadline ratio eventually increases for applications that are continually being starved. For this, it balances out the ratio by the total number of packets transmitted  to the total number of packets generated till the current time. Note that here we do not only consider the total number of packets transmitted, instead we also consider number of packets generated as they are not uniform across the stations.  
%This does not always mean that the NRLF scheduler would result in lower total packet drop profit incurred compared to the LRF scheduler, this is because I
Hence, over long run the NRLF scheduler will start prioritizing packets with lower profits over those with higher profits.

\section{Evaluation Results}
\label{sec:resultsanalysis}

\textcolor{black}{We compute all the metrics for both versions of our algorithm i.e., \lsdsf and \lsds.} We then compare it with other benchmarks i.e., EDF, LRF, and NLRF for all $4$ use cases and plot the results in Figures ~\ref{usecase1-profit}-~\ref{usecase5-profit}. Here, we observe that both \lsdsf and \lsds outperform all benchmarks for all $3$ metrics across all use cases. The primary reason for the improved performance can be attributed to the interval scheduling viewpoint which allows us to de-couple the task of selecting the correct set of transmission intervals and the correct combination of packets and RUs within each interval. Furthermore, our algorithm avoids myopic decisions unlike local greedy approaches adopted by the benchmarks. For instance, a particular job might be added and discarded by our algorithm several times till termination. On the other hand, the benchmarks naively sort the stations based on some intuitive scheduling metric and groups packets greedily. This is also why the benchmarks are much faster, but highly sub-optimal. Essentially, they fail to find the combinations which might be sub-optimal `locally' but earns a lot of benefit from a global point of view. \textcolor{black}{Further, \lsds outperforms \lsdsf as the latter only uses a fixed RU split, while the former picks the optimal RU config.}
%This might lead to gaps in the intervals ultimately chosen by these benchmarks. Hence, all the packets that could have possibly been scheduled in these gaps get dropped. In addition, \dpmss finds the optimum set of packets to be assigned to each interval whereas the simple sorting used by the benchmarks might actually miss out packets that could have been scheduled in that very interval.  That is why all the benchmarks have a higher total loss ratio compared to \dpmss. 

\subsection{Evaluation Results for Different Use Cases}

We now discuss each use case-specific result.
%\vspace*{-0.25in}

\noindent\textbf{Use Case-1:}
\begin{figure}[t]
\subfloat[]{\includegraphics[width = 0.25\textwidth]{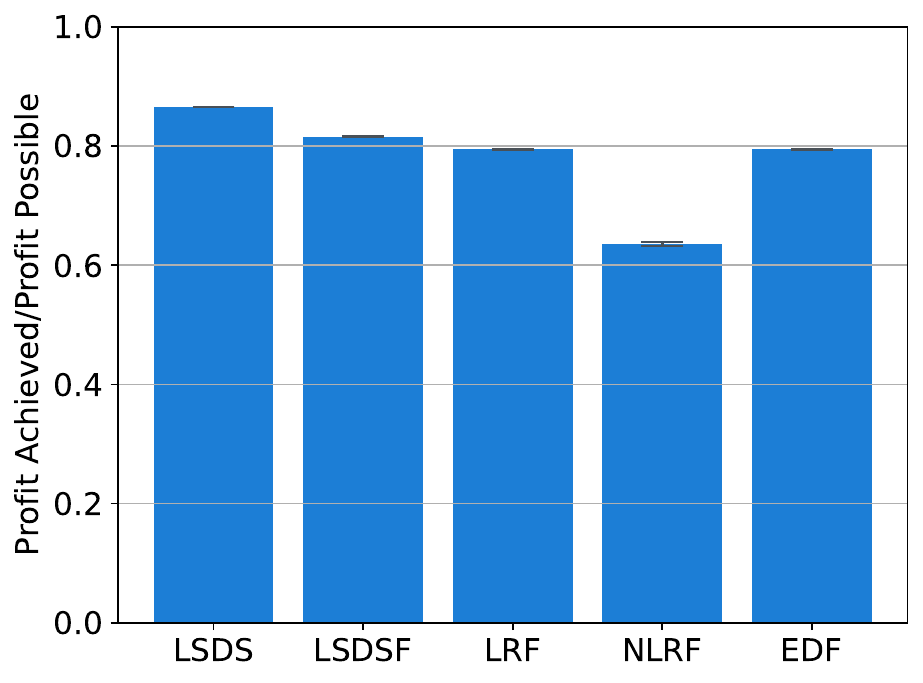}}
\subfloat[]{\includegraphics[width = 0.25\textwidth]{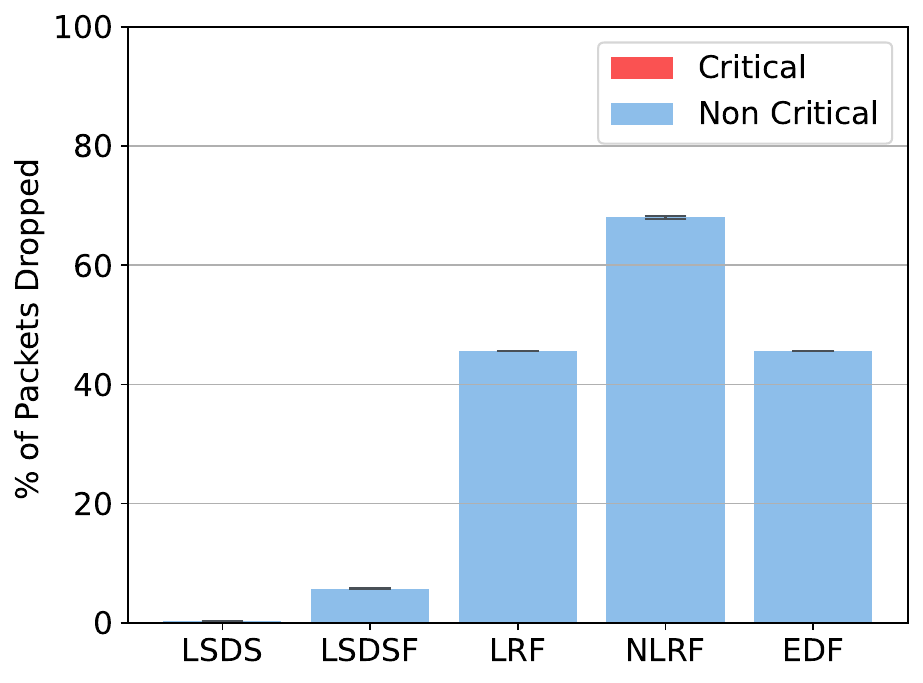}}
\vspace{-0.15in}
\caption{Use Case-1 (UC-1): LSDS, LSDSF, EDF, LRF, and NLRF; (a) Ratio of profit obtained to total profit, (b) Percentage of total packets and critical packets dropped.}
\label{usecase1-profit}
\vspace{-0.15in}
\end{figure}
Figure \ref{usecase1-profit} shows the profit-ratio and drop percentage for UC-1.  In this case, we have about $8$K packets/sec, hence even \lsds was not able to schedule all packets within their deadline. However, it outperforms all the baselines. Here, EDF and LRF also performed well and were able to obtain a profit ratio close to $0.8$.  Note that in this case, all the applications have the same profit. Hence, the scheduling metric turns out to be the same for both EDF and LRF for all applications and they behave exactly the same. However, NLRF suffers. As observed from Table.~\ref{usecase1-profit} the applications i.e., profile 1 and 2 with higher packet generation rates also have lower delay tolerance. 
 %Profile 1 and 2 have a very low delay tolerance value compared to profile 3, profile 4, and profile 5.
 This causes applications with relatively larger delay tolerance to starve and hence NLRF tries to schedule these applications. However, while scheduling packets of Profiles 3-5, NLRF misses out on Profiles 1 and 2, but they have a high generation rate with the same profit. Hence, overall NLRF achieves a lower profit compared to LRF and incurs a higher loss ratio as well.

\noindent \textbf{Use Case-2:}
\begin{figure}[b]
\vspace{-0.15in}
\subfloat[]{\includegraphics[width = 0.25\textwidth]{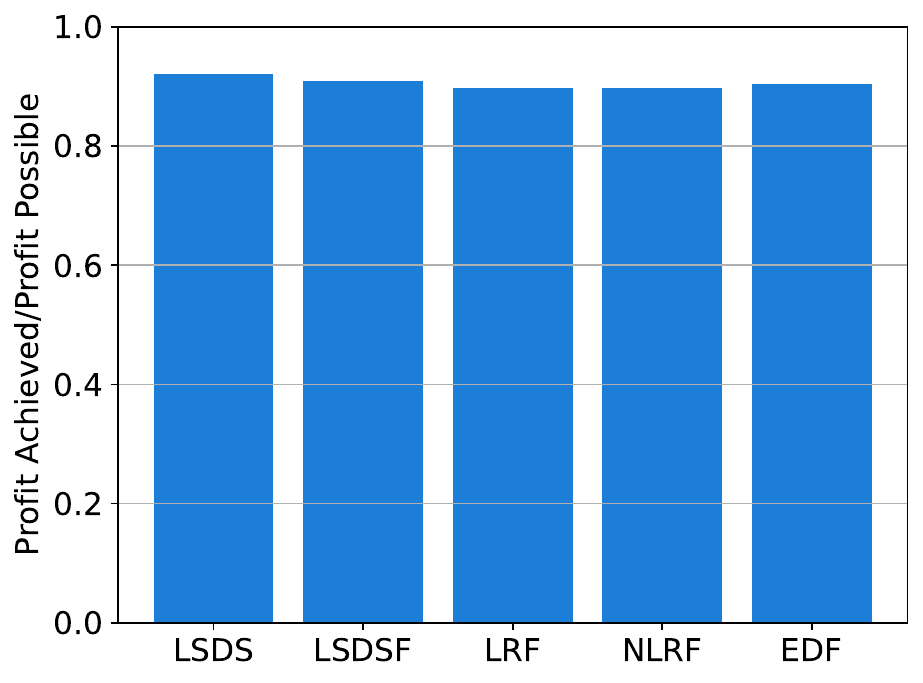}}
\subfloat[]{\includegraphics[width = 0.25\textwidth]{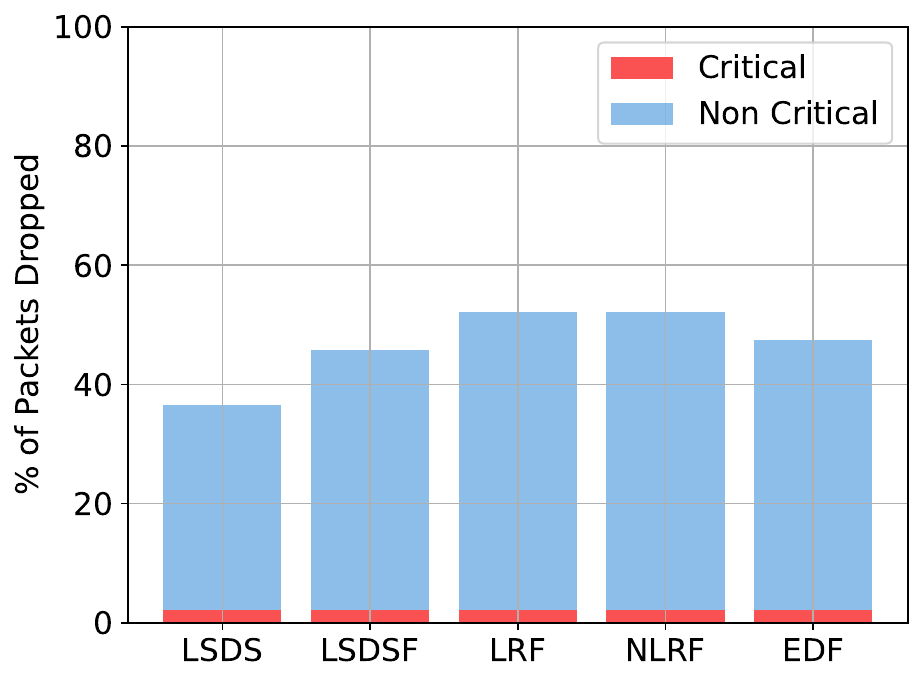}}
\vspace{-0.15in}
\caption{Use Case-2 (UC-2): LSDS, LSDSF, EDF, LRF, and NLRF; (a) Ratio of profit obtained to total profit, (b) Percentage of total packets and critical packets dropped.}
\label{usecase3-profit}
\vspace{-0.15in}
%\vspace{-0.4cm}
\end{figure}
Figure~\ref{usecase3-profit} shows the ratio for UC-2. There are a few interesting observations. The ratio for \lsds is a little less than $1$. Here, the number of packets for the most critical application i.e., the control traffic is high. It generates almost $1$ packet every ms and they have a deadline of $16$ms. However, there are $20$ nodes that run these control traffic application.  Hence, \lsds was not able to schedule them all and had to drop some critical packets too ($\simeq 2\%$). Further, video surveillance also had a very high generation rate hence causing high non-critical packet drops. 
%Further, to prioritize the critical packets, \lsds had to drop some non-critical packets, hence incurring a loss ratio of $\simeq 6\%$. All the benchmarks have a loss percentage higher than $40\%$.
Here, EDF prioritizes all $4$ applications with shorter deadlines i.e., smart-meter, status information, data polling, and control traffic. However, LRF correctly prioritizes the most critical application i.e., control traffic. However, the performance of all the algorithms is very similar as other applications like smart-meter, status information, and data polling have a very lower packet generation rate i.e., $1-2$ packets/sec. Hence, all the algorithms end up allocating resources to control and video surveillance applications only.   %In this case, NLRF incurs a very high loss rate for critical packets and an overall higher loss rate. This is because NLRF prioritizes applications with higher scheduling metric. Here, applications like reporting and logging, and video surveillance suffer due to their higher deadline and relatively lesser profit. Hence, NLRF prioritizes them and hence incurs a drop of packets from critical applications. Hence, overall it incurs a lesser profit ratio as well. 

\noindent \textbf{Use Case-3:}
%\vspace{-0.15in}
\begin{figure}[t]
%\vspace{-0.5cm}
\subfloat[]{\includegraphics[width = 0.25\textwidth]{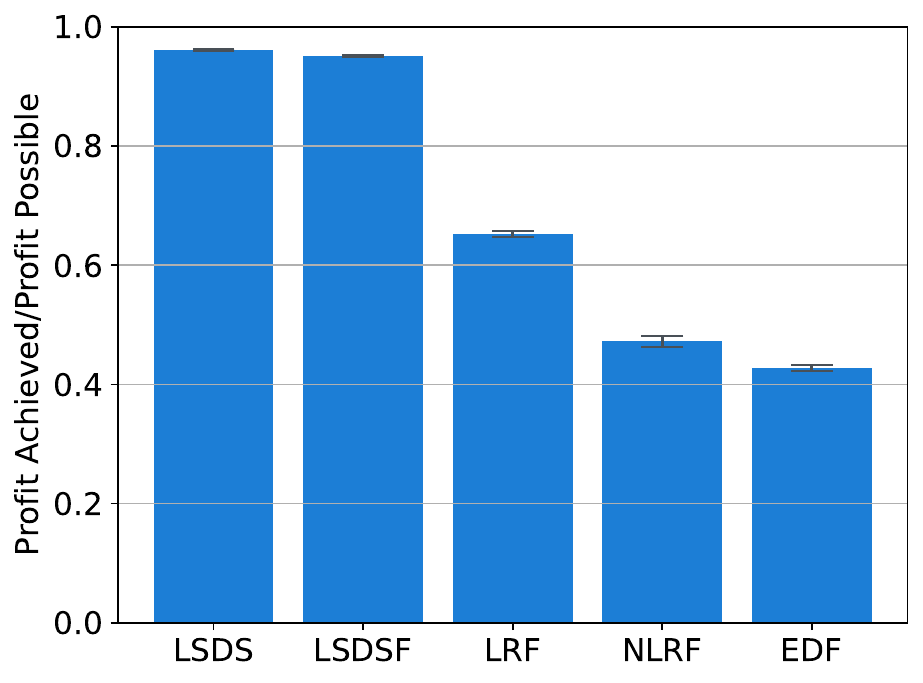}}
\subfloat[]{\includegraphics[width = 0.25\textwidth]{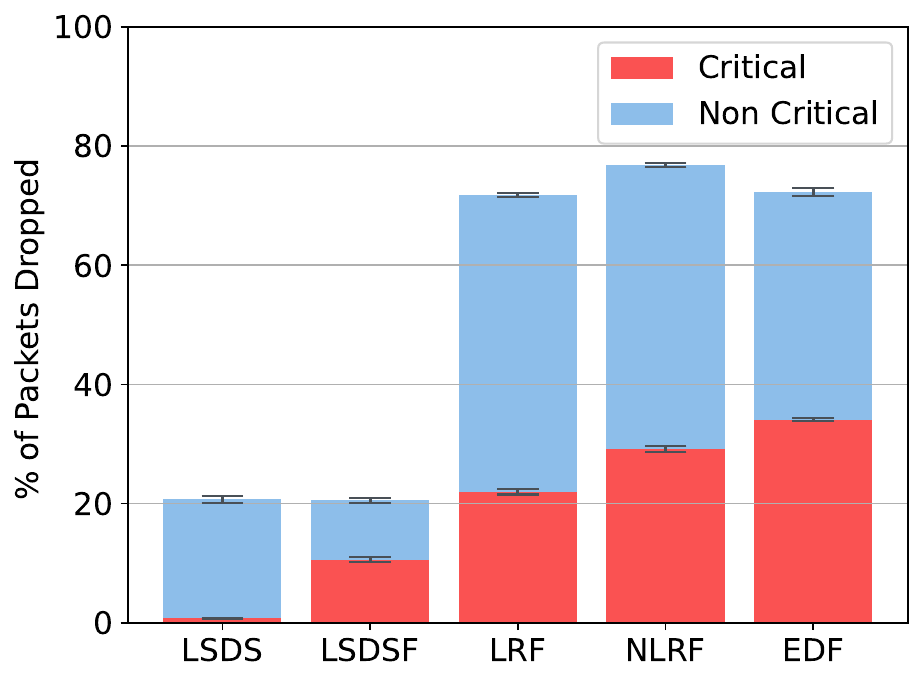}}
\vspace{-0.15in}
\caption{Use Case-3 (UC-3): LSDS, LSDSF, EDF, LRF, and NLRF; (a) Ratio of profit obtained to total profit, (b) Percentage of total packets and critical packets dropped.}
\vspace{-0.15in}    
\label{usecase4-profit}
%\vspace{-0.4cm}
\end{figure}
Figure~\ref{usecase4-profit} shows the ratio for UC-3. The total capacity requirement for this use case is $16,00000\times50\times8=640Mbps$. A bandwidth of $40$ MHz cannot handle this much load (as the maximum bitrate is about $510$ Mbps at MCS11), hence, we run this use case for higher bandwidth of $160$MHz.  Interestingly, here \lsdsf and \lsds have similar packet drop percentages, however, the number of critical packet drops is greater for \lsdsf.  Unlike \lsdsf, \lsds not only generates optimum scheduling intervals but also finds out the optimum RU split to be used for the packet-to-interval mapping. This results in  lower critical drops in \lsds than \lsdsf. Notably, here EDF performs poorly as it prioritizes the applications with lower deadlines i.e., motion control and collaborative AGV. It misses out on robotic control that has a very high profit. Hence, it obtains a poor profit overall. Here, LRF correctly prioritizes all $3$ important applications i.e, motion control, collaborative AGV, and robotic control. Hence, it obtains overall higher profit compared to EDF. Further, the critical loss ratio for LRF is lower compared to EDF. However, this lower ratio comes with a trade-off of higher overall loss rate for non-critical packets. Since this causes starvation for asset/process monitoring application, NLRF prioritizes that. But, due to that, it misses packets from all the critical applications as the asset/process monitoring application as well has a very high generation rate. Hence, overall NLRF obtains a  poor profit ratio and  high loss percentage.  %Interestingly, in this usecase two applications who have a small deadline have a high profit too. Hence, EDF performs well. Similarly, for LRF the ratio between profit/deadline is high for motion control, collaborative AGV hence prioritizes them. This causes starvation for robotic control and asset/process monitoring application. However, all the applications have the same transmission rate. Hence, 

\noindent \textbf{Use Case-4:}
\begin{figure}[t]
\vspace{-0.12in}
\subfloat[]{\includegraphics[width = 0.23\textwidth]{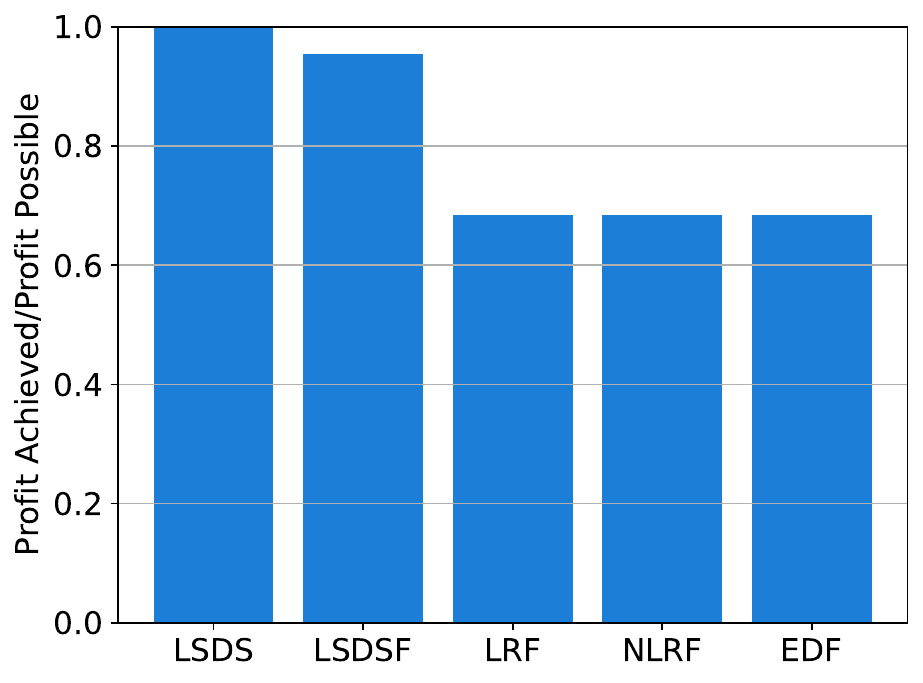}}
\subfloat[]{\includegraphics[width = 0.23\textwidth]{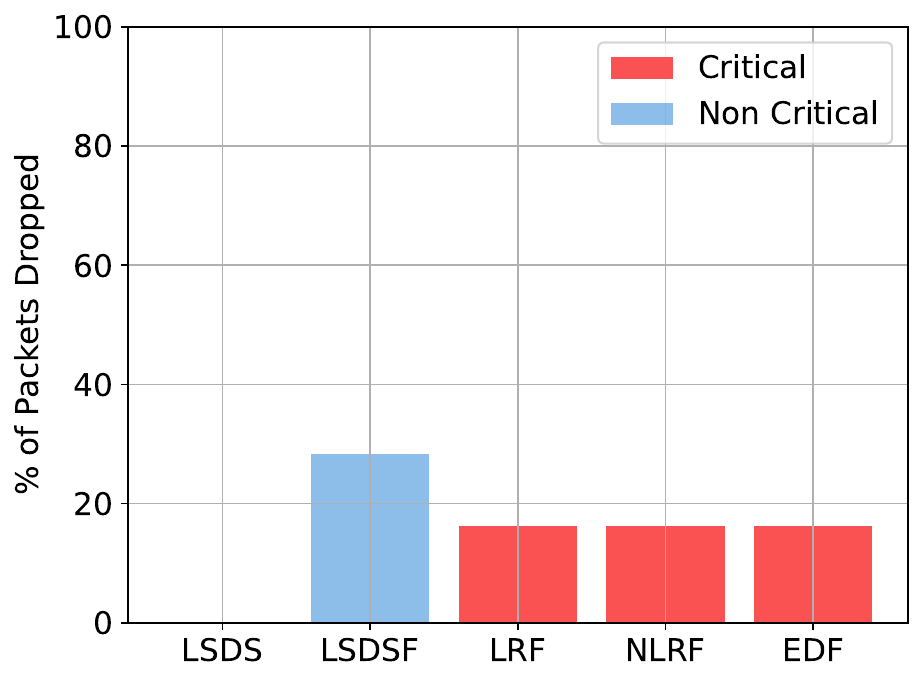}}
\vspace{-0.15in}
\caption{Use Case-4 (UC-4): LSDS, LSDSF, EDF, LRF, and NLRF; (a) Ratio of profit obtained to total profit, (b) Percentage of total packets and critical packets dropped.}
\label{usecase5-profit}
\vspace{-0.15in}
\end{figure}
Figure~\ref{usecase5-profit} shows the ratio for UC-4. Here, we have a very interesting observation. \lsds incurs $0\%$ packets drops and hence a profit ratio of $1$. Even the fixed split version \lsdsf incurs $0\%$ critical drop percentage and hence a profit ratio very close to 1. The other three benchmarks incur about $18\%$ critical packet drop percent. The reason is, EDF prioritizes application $6$ and then application $2$ for their smaller deadline. Interestingly, in this use case, two applications that have a small deadline have a high profit too. Hence, both EDF and LRF perform the same. Here, the starvation will be for applications $1$, $3$, $7$, $10$ as their scheduling metric is very less. Hence, NLRF prioritizes them. But, the generation rate of these applications is very small, atmost $1$ packets/sec for application $1$, $3$, and $7$. Though the generation rate for application $10$ is relatively higher, it has only $1$ node and the delay tolerance value is really high. Hence, the effect of scheduling these applications is minimal. Hence, all the benchmarks perform very similarly. Now, EDF prioritizes application $6$ and $2$ and application $6$ has a very low generation rate. As a result, the schedule created by EDF will have large gaps, that will cause another critical applications i.e., application $4$ to drop packets. The same reasoning is true for LRF too as for LRF the scheduling metric is highest for application $6$ followed by application $2$. The drop is caused by application $4$ which is critical. Hence, all the benchmarks incur $\simeq 18\%$ critical packet drops. We note here that both \lsds and \lsdsf, by design prioritize packets that are of significantly high profit and schedules them in the appropriate  interval. Hence, both obtain a higher profit ratio with no critical packet drops. Finally, several noncritical applications have higher packet sizes i.e., applications $1, 7, 10$. \lsdsf was not able to schedule them in small $26$ tone RUs within TXOP duration, and hence incurs noncritical packet drops. 
%all the non critical packets have a very high delay tolerance too. Hence, all the algorithms were able to schedule them within their due time without dropping.

These results show that \lsds outperforms all the other baselines in each case. While \lsdsf performs worse than \lsds due to its fixed configuration of RU's, it still outperforms LRF, NLRF and EDF. This shows that both local search to schedule packets optimally (used in both \lsdsf and \lsds) and search of RU configuration (used only in \lsds) significantly improve the solution to the \dpmss problem.

\subsection{Runtime Analysis} 
\begin{figure}[t]
    \centering
    %\vspace{-0.1in}
    \includegraphics[width=0.24\textwidth]{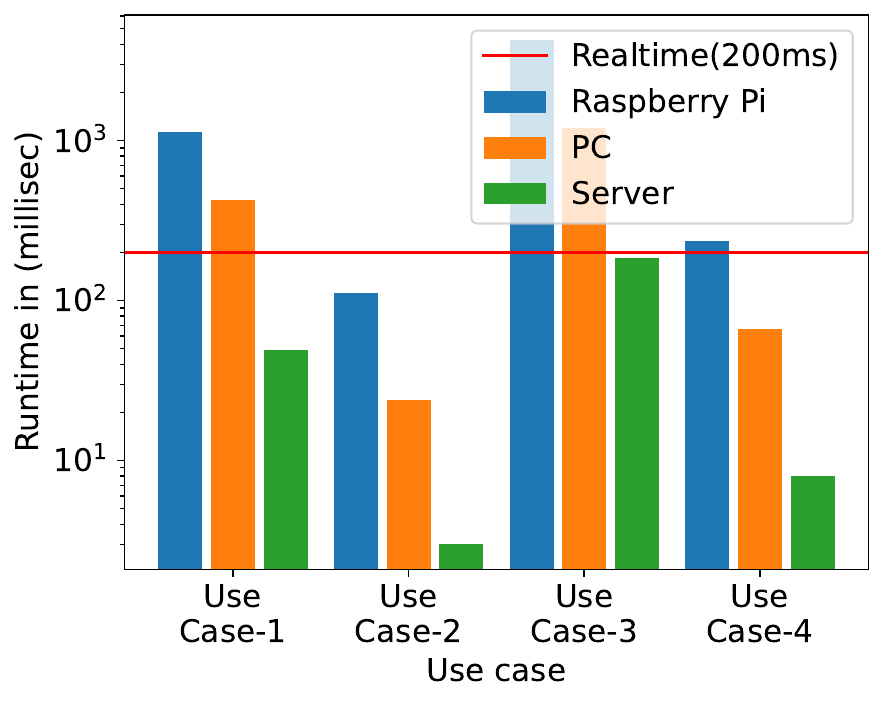}
   \vspace{-0.2in}
    \caption{Runtime of \lsds on three different platforms.}
    \vspace{-0.15in}
    \label{fig:my_label}
\end{figure}
We now plot the runtimes of our \lsds scheduler on three different platforms -- a Raspberry Pi 4 having  Quad core Cortex-A72 of frequency 1.5GHz, an AMD Ryzen 7 laptop with maximum frequency 3.2 GHz and an Intel Core i7-11700B processor server with maximum frequency 3.2 GHz. We omit the benchmarks since their metrics are trivial to compute. We have different time horizons in each case, in each case smaller than $200$ ms. We note that for two of the use cases UC-2 and UC-4, the execution time is smaller than $200$ ms across all platforms. For UC-1 and UC-3, the runtime is smaller than $200$ ms on the server. 
%Only for UC-4, the runtime exceeds $200$ ms. 
This indicates that the \lsds scheduler can run in real-time for all cases on a server. 
Even on less compute-intensive systems, the easier use cases (such as UC-2) run in real-time, which makes integration of \lsds less expensive in such situations.
%but the most challenging workloads in real-time. 
%Even for UC-4, the execution time on server is around $1$ s, which is still usable in practice.

\subsection{Performance Under Poor Channel Condition}
\begin{figure}[t]
\centering
\vspace{-0.1in}
\subfloat[]{\includegraphics[width = 0.22\textwidth]{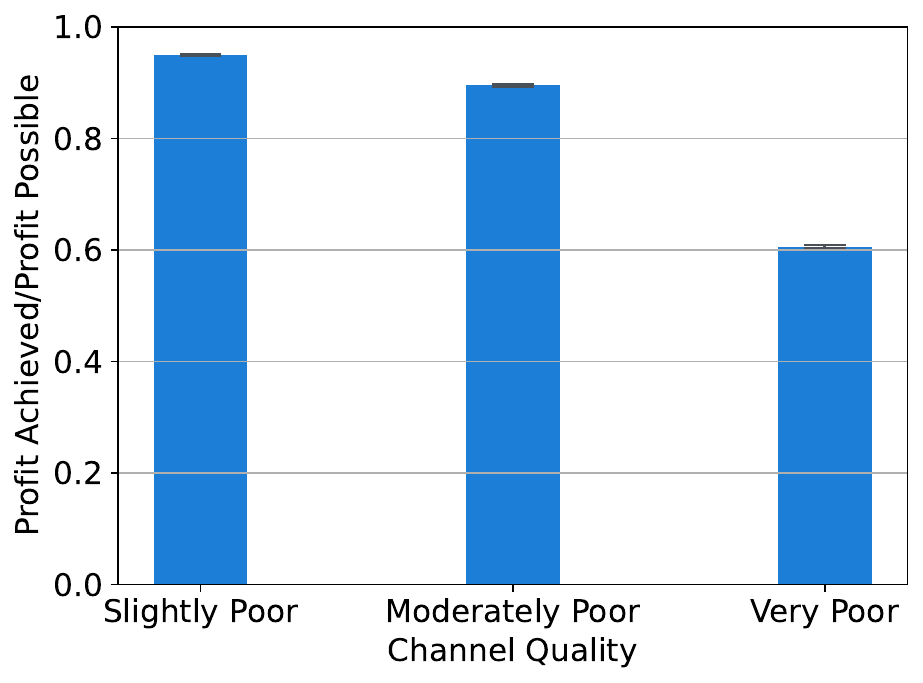}}
\subfloat[]{\includegraphics[width = 0.22\textwidth]{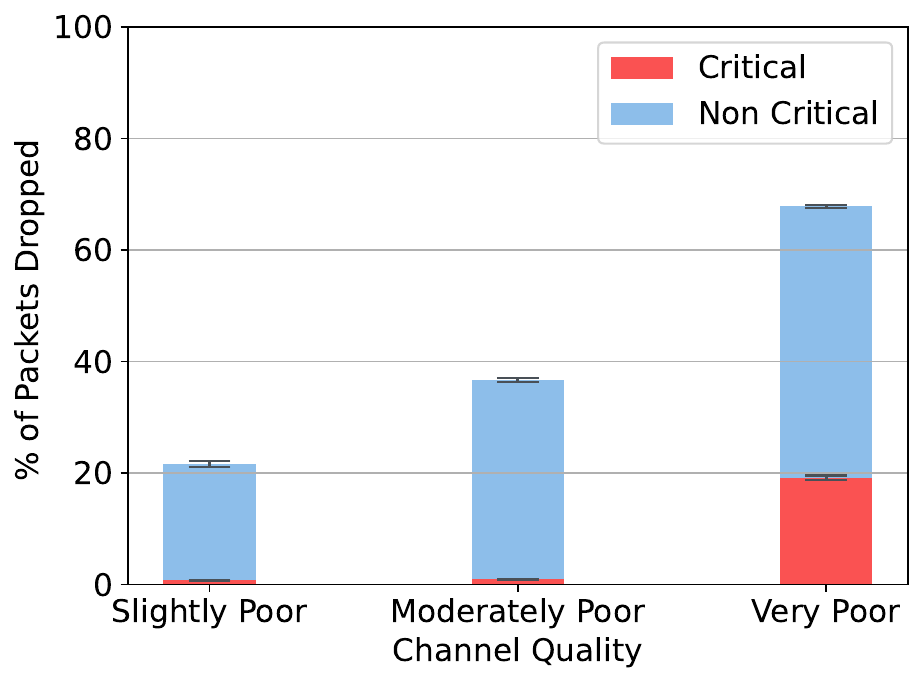}}
\vspace{-0.2in}
\caption{Performance of LSDS under poor channel in terms of (a) Ratio of profit obtained to
total profit, and (b) Percentage of total packets and critical packets
dropped.}
\label{usecase4-poor}
\vspace{-0.15in}
\end{figure}

We now show that LSDS performs well in more challenging conditions, where path losses can lead to lower throughput.
To do so, we run the simulations for three different scenarios where the clients are distributed from $5$ m, $15$ m, and $40$ m radius from the AP. We label these conditions as slightly poor, moderately poor, and very poor respectively.
We follow IEEE 802.11ax path loss model~\cite{path-loss}, thus correspondingly reducing the MCS. To make this discussion brief, we present the results only for the most critical use case i.e., UC-3.
Here, we observe that no critical packets are dropped by \lsds in slightly poor and moderately poor conditions. Only in very poor channel conditions, \lsds drops critical packets. The reason for the same is that with poorer channel the MCS drops and hence the transmission time of frames increase. Now, for UC-3 there is $16,00000$ packets/sec, now as the transmission time of each individual OFDMA frame increases, not all packets could get scheduled by their deadline. 

\iffalse
\subsection{Interference}
Next, we consider a dense deployment of multiple BSSs where the APs operate on overlapping channels. Hence, they interfere with each other. Here, we consider frequency selective fading where some of the RUs suffer from interference. Here again, we consider 3 cases where the number of RUs facing interference is less than less $xx$\%, moderate i.e., $xx$\%, and high i.e., $xx$\%. 
We show the results for UC-4.
\fi
\subsection{Support for Best-effort Traffic}
\begin{figure}[b]
\centering
\vspace{-0.2in}
\subfloat[]{\includegraphics[width = 0.23\textwidth]{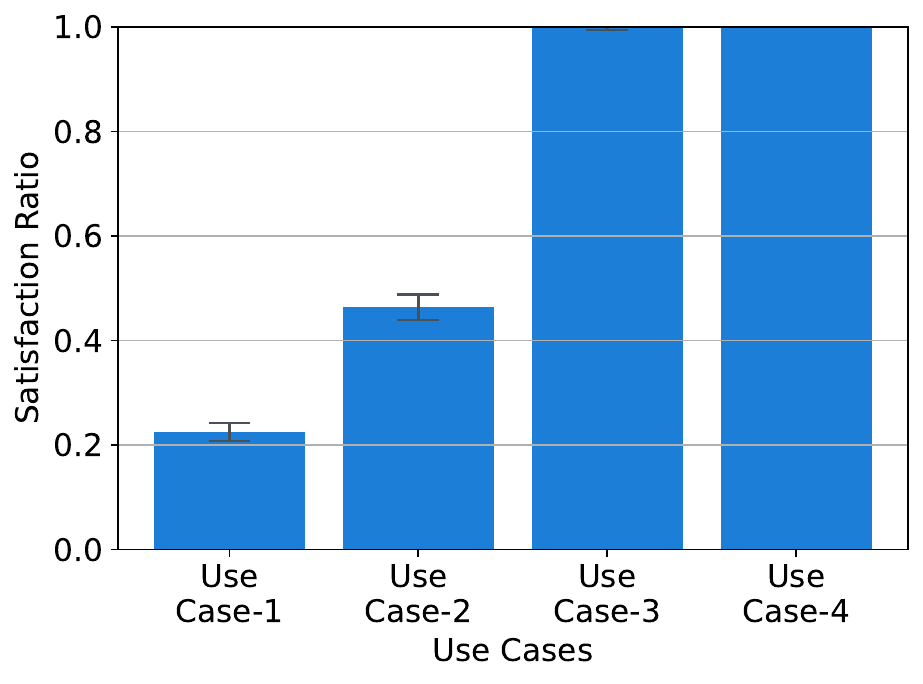}}
\subfloat[]{\includegraphics[width = 0.23\textwidth]{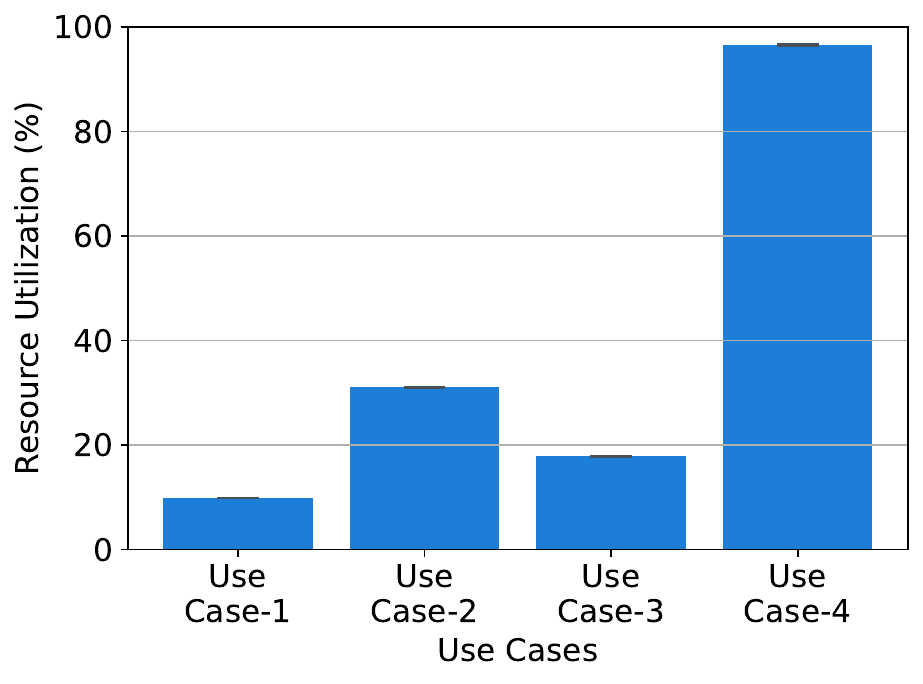}}
\vspace{-0.2in}
\caption{Support for best-effort traffic: satisfaction ratio and resource utilization}
\label{usecase5-profit}
\vspace{-0.2in}
\end{figure}

In factory settings, apart from the factory-based applications, the same WiFi network can be used by supporting best-effort Internet applications used by the factory workers such as web browsing, video streaming, and so on. Hence, we propose a mechanism where the network is shared by both factory applications and best-effort ones. 
If a best-effort packet arrives at the $i^{th}$ round, we try to schedule it into either the $i^{th}$ or $(i+1)^{th}$ round's free RUs i.e., after allocating RUs to factory applications. If the packet does not get scheduled in either $i^{th}$ or $(i+1)^{th}$ round, we convert it into a deterministic packet for
the $(i+1)^{th}$ round and continue the process. Also, to make sure the packet gets
scheduled eventually, the profit is increased after every round as follows:
$p_j^i = (p_j^{i-1} +criticalThreshold)/2$ where $p_j$ is the profit of packet $p$.
In addition to assigning best-effort traffic to free resources in the frequency domain, we utilize free time slots that are left unoccupied after assigning factory applications.

We show the satisfaction ratio for best-effort applications along with resource utilization solely by best-effort traffic. We define the satisfaction ratio as the throughput obtained to the offered load.
We create $3$ best-effort nodes and they generate a load following Poisson distribution with a mean of $20Mbps$ for UC-1, 2, and 4.  Since UC-3 was run at $160$ Mhz, the mean is $80$ Mbps.
We observe that for both UC-1 and UC-2, there are not many free RUs and hence the satisfaction ratio is about $0.2$ and $0.5$ respectively. 
We observe that for UC-4, best-effort traffic utilizes close to $90$\% resources and obtain a satisfaction ratio of $1$. This corroborates our previous observation that \lsds was not dropping any packets for this use case. This indicates UC-4 did not have high traffic load and hence most RUs were free.  Since UC-3 was run at $160$ MHz, even though the resource utilization by best-effort traffic is less than $20$\%, it was easily able to handle a load of $80$Mbps. Hence, it provides a satisfaction ratio of $1$. This shows that our technique of scheduling packets can efficiently accommodate additional best-effort traffic if it is necessary.

\section{Conclusion}
Integration of wireless networks in factory environments has been long overdue, and WiFi standards have been taking concrete steps in this direction. 
In line with these efforts, our work models the scheduling requirements using a system of packets and deadlines for each data packet in such scenarios. 
We then provide a deadline-aware scheduler for factory IoT settings. We show that the problem is NP-Hard and provide a $(12+\varepsilon)$-approximation algorithm using a technique of local search. We introduce a novel variant of deadline scheduling which we hope would spark independent interest in the scheduling community. We evaluate the effectiveness of our algorithm through various simulations in terms of number of packets delivered, running time of our algorithm as well as adaptation to poor channel conditions and additional network load. We observe that our algorithm works well in practice and outperforms all the benchmarks. We believe this work provides a reason for the consideration of WiFi as a feasible technology in the industrial space. For future work, we plan to consider more dynamic situations where the deadlines and the packet arrivals are stochastic in nature.
%\newpage
\bibliographystyle{plain}
%\vspace{-0.05cm}
\bibliography{reference}

\end{document}